%% file: main.tex
\documentclass[a4paper]{article}

\usepackage[english]{babel}
\usepackage[utf8]{inputenc}

\usepackage[round]{natbib}
\setcitestyle{authoryear}

\usepackage{amsmath}
\usepackage{graphicx}
\usepackage{amsfonts}
\usepackage{tikz}
\usetikzlibrary{shapes,arrows}
\usetikzlibrary{backgrounds}
\usetikzlibrary{positioning}
\usepackage{listings}
\usepackage{color}
\usepackage[margin=1in]{geometry}

\definecolor{mygreen}{rgb}{0,0.6,0}
\definecolor{mygray}{rgb}{0.5,0.5,0.5}
\definecolor{mymauve}{rgb}{0.58,0,0.82}

\lstdefinelanguage{scala}{
  morekeywords={abstract,case,catch,class,def,%
    do,else,extends,false,final,finally,%
    for,if,implicit,import,match,mixin,%
    new,null,object,override,package,%
    private,protected,requires,return,sealed,%
    super,this,throw,trait,true,try,%
    type,val,var,while,with,yield},
  otherkeywords={=>,<-,<\%,<:,>:,\#,@},
  sensitive=true,
  morecomment=[l]{//},
  morecomment=[n]{/*}{*/},
  morestring=[b]",
  morestring=[b]',
  morestring=[b]"""
}

\newcommand\numberthis{\addtocounter{equation}{1}\tag{\theequation}}

\title{Composable Models for Online Bayesian Analysis of Streaming Data}

\author{J. Law, \quad D. J. Wilkinson\\ School of Mathematics \& Statistics\\Newcastle University, U.K.}

\date{\today}

\begin{document}
\maketitle

\begin{abstract}
Data is rapidly increasing in volume and velocity and the Internet of Things (IoT) is one important source of this data. The IoT is a collection of connected devices (things) which are constantly recording data from their surroundings using on-board sensors. These devices can record and stream data to the cloud at a very high rate, leading to high storage and analysis costs. In order to ameliorate these costs, we can analyse the data as it arrives in a stream to learn about the underlying process, perform interpolation and smoothing and make forecasts and predictions.

Conventional tools of state space modelling assume data on a fixed regular time grid. However, many sensors change their sampling frequency, sometimes adaptively, or get interrupted and re-started out of sync with the previous sampling grid, or just generate event data at irregular times. It is therefore desirable to model the system as a partially and irregularly observed Markov process which evolves in continuous time. Both the process and the observation model are potentially non-linear. Particle filters therefore represent the simplest approach to online analysis. A functional Scala library of composable continuous time Markov process models has been developed in order to model the wide variety of data captured in the IoT.
\end{abstract}

\section{Introduction}
\label{sec:intro}
Observations across sensor networks are highly heterogeneous, such as weather and temperature, counts of cars and bikes or tweets from Twitter. In order to model this data we need a wide variety of observation models, not just the Normal distribution. Dynamic Linear Models (DLMs) are often applied to modelling time series data. They are discrete time, latent variable models which can be applied to a wide variety of problems, provided the data is Normally distributed and the model is linear. The Kalman filter is an analytical solution to find the distribution of the latent variables of a DLM, which can be used to perform forecasting~\citep{Kalman1960}. The closed form solution can be found in the case of a linear-Gaussian model because of special properties of the Gaussian distribution; the sum of two Gaussian distributions is a Gaussian distribution and a linear translation of a Gaussian distribution is still Gaussian with predictable mean and variance.

In order to model non-linear systems, the extended (EKF) and later unscented Kalman filter (UKF) have been developed~\citep{Julier1997}. The extended Kalman filter linearises at the current time step, by calculating the Jacobian of the transition and observation functions at the given time and using these in the Kalman filter equations. The extended Kalman filter becomes unstable when applied to highly non-linear problems, so the unscented transform was introduced. 

Particle filters can determine the state space of a non-linear, non-Gaussian latent variable model, with minimal modification between models. The UKF is more computationally efficient than the simulation based particle filter and can be more accurate when models are near-linear~\citep{Bellotto2007}. However, particle filters allow more flexibility of model choice and as such are considered when performing inference for the composable models presented in this paper. As the number of particles in the particle filter is increased, the estimation error tends to zero, the same is not true of the UKF, since a limited number of sigma points are used~\citep{simon2006optimal}. So with access to more computing power, the particle filter is preferred for accurate inference.

\cite{Gordon1993} developed the theory of the bootstrap particle filter and compared the performance of the new filter to the Extended Kalman Filter (EKF). They found the performance of the bootstrap filter to be greatly superior to the EKF using a highly non-linear model. A target tracking model was also considered, where the bootstrap filter again outperforms the EKF.

Partially observed Markov processes~\citep{Ionides2006} with irregular observations are used in order to model the variety of data streaming from sensor networks in the internet of things. The evolution of the unobserved system state is governed by a diffusion process, a continuous-time Markov process. A diffusion process is the solution to a stochastic differential equation.  The Markov property of the diffusion process can be used to forward simulate the state between irregular observations given the previous state estimate and time different between observations.

POMP models are flexible in the choice of state distribution and observation distribution, hence they can be used to model a wide variety of processes, such as counting processes. For a full Bayesian analysis of a POMP model, the full-joint posterior distribution of these models can be determined using the Particle Marginal Metropolis Hastings algorithm (PMMH)~\citep{Andrieu2010}. The PMMH algorithm requires a batch of observations to determine the parameters. On-line filtering and forecasting can be carried out using a particle filter with per-determined parameters, and occasional retraining of the model can be carried out as required. On-line learning of model parameters is an active research area not pursued here; see~\cite{Carvalho2010} for more details.

It is useful to reuse and combine existing models when exploring new data and building more complex models. The POMP models considered in this paper can be composed together to create new models. A software package has been written in Scala~\citep{Odersky2004} which allows users to build and compose models and perform inference using the bootstrap particle filter and the PMMH algorithm.

There are other software packages for performing inference on POMP models: LibBi \citep{Murray2015} implements inference for general state-space models using particle Monte Carlo Markov chain (pMCMC) methods and sequential Monte Carlo (SMC). It is optimised for parallel hardware and utilises CUDA (a parallel programming platform by NVIDIA) for GPU (Graphics Processing Unit) programming. There is also an R~\citep{rproject} package, POMP~\citep{King2016} which implements Frequentist and Bayesian methods of parameter inference for POMP models. However, neither package supports online analysis of streaming data.

The composable POMP models presented in this paper have been developed to analyse data in the Urban Observatory~\citep{uo2014Temperature}, a grid of sensors deployed around the city of Newcastle Upon Tyne. In order to demonstrate the utility of this class of composable POMP models, there are two examples presented in Section~\ref{sec:example}. The first example consists of arrival times of traffic at specific locations in the city, which is modelled using a Poisson distribution with seasonally varying rate. The second example consists of temperature data collected from urban sensors in Newcastle Upon Tyne~\citep{uo2014Temperature} and demonstrates the application of continuous time state space POMP models to irregularly observed time series data.

Section~\ref{sec:POMP} introduces Partially Observed Markov Processes for modelling a variety of time evolving data. This includes binary (Bernoulli), count (Poisson), time to event (Log-Gaussian Cox-Process) and seasonal data. Section~\ref{sec:composed} describes how to compose models together to enable complex models to be developed. Section~\ref{sec:bootstrap} presents the bootstrap particle filter for calculating the latent state of the model, and how it can be used to calculate an estimate of the marginal likelihood of the observations given a set of parameters. Section~\ref{sec:pmmh} outlines the Particle Marginal Metropolis Hastings algorithm, which can be used to calculate the full joint posterior of a POMP model.

\section{Partially Observed Markov Processes}
\label{sec:POMP}

Streaming data arrives frequently and discretely as an observation, $y$, and a timestamp, $t$. In order to model this time series data, a structured model is proposed:

\begin{align*}
	Y(t_i)|\eta(t_i) &\sim \pi(y(t_i) | \eta(t_i), \theta), \\
  	\eta(t_i)|\textbf{x}(t_i) &= g (F_{t_i}^T\textbf{x}(t_i)), \\
	\textbf{X}(t_i)|\textbf{x}(t_{i-1}) &\sim p(\textbf{x}(t_i)|\textbf{x}(t_{i-1}), \theta), \quad \textbf{x}(t_0) \sim p(\textbf{x}(t_0) | \theta). \numberthis \label{eqn:genpomp}
\end{align*}

$Y(t_i)$ denotes the observation at time $t_i$, where $\pi(y(t_i)|\eta(t_i), \theta)$ is the observation distribution. $\textbf{X}(t_i)$ is the (possibly multidimensional) unobserved state of the process. The state is Markovian and is governed by the transition kernel $p(\textbf{x}(t_i)|\textbf{x}(t_{i-1}), \theta)$ that we assume realisations can be generated from, but in general may not represent a tractable distribution. The prior distribution on the state is given by $p(\textbf{x}(t_0) | \theta)$. $F_t$ is a time dependent vector, the application results in $\gamma(t) = F_t^T \textbf{x}(t)$, where $\gamma(t) \in \mathbb{R}$. The function $g: \mathbb{R} \rightarrow \mathcal{D}$ is a link function allowing a non-linear transformation as required by the observation distribution.

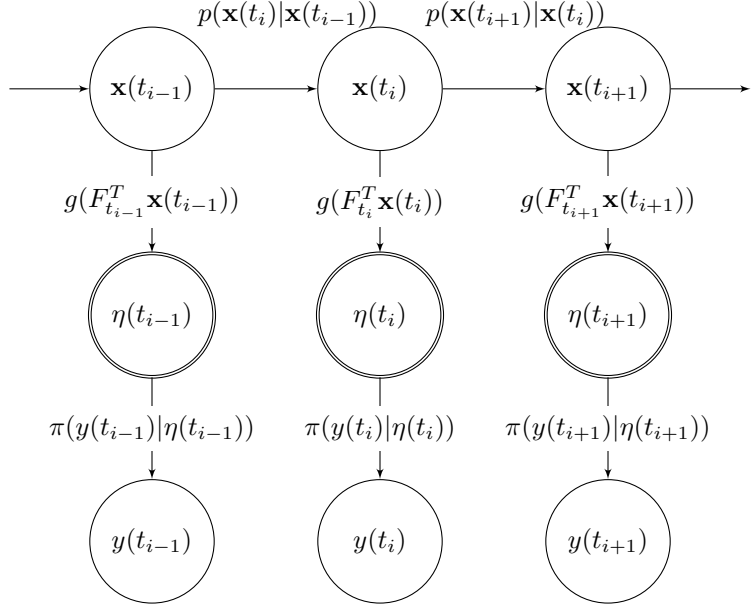
\begin{figure}
\tikzstyle{line} = [draw, -latex']
\tikzstyle{state} = [draw, circle, text width=1.3cm, align=center]
\tikzstyle{detstate} = [draw, double, circle, text width=1.3cm, align=center]

\begin{center} 
\begin{tikzpicture}[node distance = 2cm, auto]
    \node (1) {};
    \node [state, right of=1] (2) {$\textbf{x}(t_{i-1})$};
    \node [state, right of=2, node distance=3cm] (3) {$\textbf{x}(t_i)$};
    \node [state, right of=3, node distance=3cm] (4) {$\textbf{x}(t_{i+1})$};
    \node [right of=4, align=center, node distance=2cm] (5) {};
    \node [detstate, below of=2,node distance=3cm] (6) {$\eta(t_{i-1})$};
    \node [detstate, below of=3,node distance=3cm] (7) {$\eta(t_{i})$};
    \node [detstate, below of=4, node distance=3cm] (8) {$\eta(t_{i+1})$};
    \node [state, below of=6, node distance=3cm] (9) {$y(t_{i-1})$};
    \node [state, below of=7, node distance=3cm] (10) {$y(t_i)$};
    \node [state, below of=8, node distance=3cm] (11) {$y(t_{i+1})$};
    
    \path [line] (2) -- (3);
    \path [line] (3) -- (4);
    \path [line] (1) -- (2);
    \path [line] (4) -- (5);
    \path [line] (2) -- (6);
    \path [line] (3) -- (7);
    \path [line] (4) -- (8);
    \path [line] (6) -- (9);
    \path [line] (7) -- (10);
    \path [line] (8) -- (11);
    
    \node [above right=0.1cm and -0.1cm of 2, node distance=1.5cm] {$p(\textbf{x}(t_i)|\textbf{x}(t_{i-1}))$};
    \node [above right=0.1cm and -0.1cm of 3, node distance=1.5cm] {$p(\textbf{x}(t_{i+1})|\textbf{x}(t_i))$};
    \node [rectangle, below of=2, node distance=1.5cm, fill=white] {$g(F_{t_{i-1}}^T \textbf{x}(t_{i-1}))$};
    \node [rectangle, below of=3, node distance=1.5cm, fill=white] {$g(F_{t_i}^T \textbf{x}(t_i))$};
    \node [rectangle, below of=4, node distance=1.5cm, fill=white] {$g(F_{t_{i+1}}^T \textbf{x}(t_{i+1}))$};
    \node [rectangle, below of=6, node distance=1.5cm, fill=white] {$\pi(y(t_{i-1})|\eta(t_{i-1}))$};
    \node [rectangle, below of=7, node distance=1.5cm, fill=white] {$\pi(y(t_i)|\eta(t_{i}))$};
    \node [rectangle, below of=8, node distance=1.5cm, fill=white] {$\pi(y(t_{i+1})|\eta(t_{i+1}))$};
    
\end{tikzpicture}
\end{center}
\caption{Representation of a POMP model as a Directed Acyclic Graph (DAG) \label{fig:dag}}
\end{figure}

The model is expressed as a directed graph in Figure~\ref{fig:dag}. Although the latent state evolves continuously, it is indicated only when there is a corresponding observation of the process. 

\subsection{Exponential Family Models}

Not all processes have Gaussian observation models. To this end, we propose modelling observations using exponential family models which include, Poisson, Bernoulli, Gamma, Exponential and the Gaussian distribution among others.

The exponential family of dynamic models are parameterised by $x(t_i)$, $V(t_i)$ and three known functions $b(Y(t_i), V)$, $S(Y(t_i))$, $a(x(t_i))$, the density is given by:

\begin{equation}
p(Y(t_i)|x(t_i), V) = b(Y(t_i), V) \exp \left \{\frac{S(Y(t_i))x(t_i) -a(x(t_i))}{V} \right\},
\label{eqn:expgen}
\end{equation}

$x(t_i)$ is called the natural parameter of the distribution, $V > 0$ is the scale parameter, $S(x(t_i))$ is the sufficient statistic, $b(Y(t_i), V)$ is the base measure and $a(x(t_i))$ is known as the log-partition. Our model class is then a generalisation of dynamic generalised linear models~\citep{West1997}.

\subsubsection{Binary Outcomes: The Bernoulli Distribution}

The Bernoulli distribution is used to model a binary outcome. It is parameterised by $p$, representing the probability of success (or true). In order to represent observations from a Bernoulli distribution within the exponential family of distributions the natural parameter is the logit function, $x(t_i) = \ln \frac{p(t_i)}{1-p(t_i)}$, $b(Y(t_i), V) = 1$, $S(Y(t_i)) = Y(t_i)$, $a(x(t_i)) = \ln (1 + e^{x(t_i)})$ and $V = 1$. The realm of $Y(t_i)$ is $\mathcal{Y}_t = \{1, 0\}$, which gives a value for the probability mass function of:

\begin{equation*}
\pi(Y(t_i)|p(t_i)) = \begin{cases} p(t_i) & \text{if }Y(t_i)=1, \\[6pt]
1-p(t_i) & \text {if }Y(t_i)=0.\end{cases}
\end{equation*}

The natural parameter $x(t_i)$, is the unobserved latent state and varies in continuous time according to a Markov process. The logistic function is the linking function used to transform $x(t_i)$ into a value suitable to represent a probability, $p(t_i) = \frac{\exp(x(t_i))}{1+\exp({x(t_i)})}$. A model specification with a generalised Brownian motion state (see Section~\ref{sec:state}) is presented below:

\begin{align*}
\pi(Y(t_i)|p(t_i)) &= \textnormal{Bernoulli}(p(t_i)), \\
p(t_i)|x(t_i) &= \frac{\exp(x(t_i))}{1+\exp(x(t_i))}, \\
\textrm{d}X(t_i) &= \mu \textrm{d}t + \sigma \textrm{d}W(t_i).
\end{align*}

\begin{figure}
\centering
\includegraphics[width=0.5\textwidth]{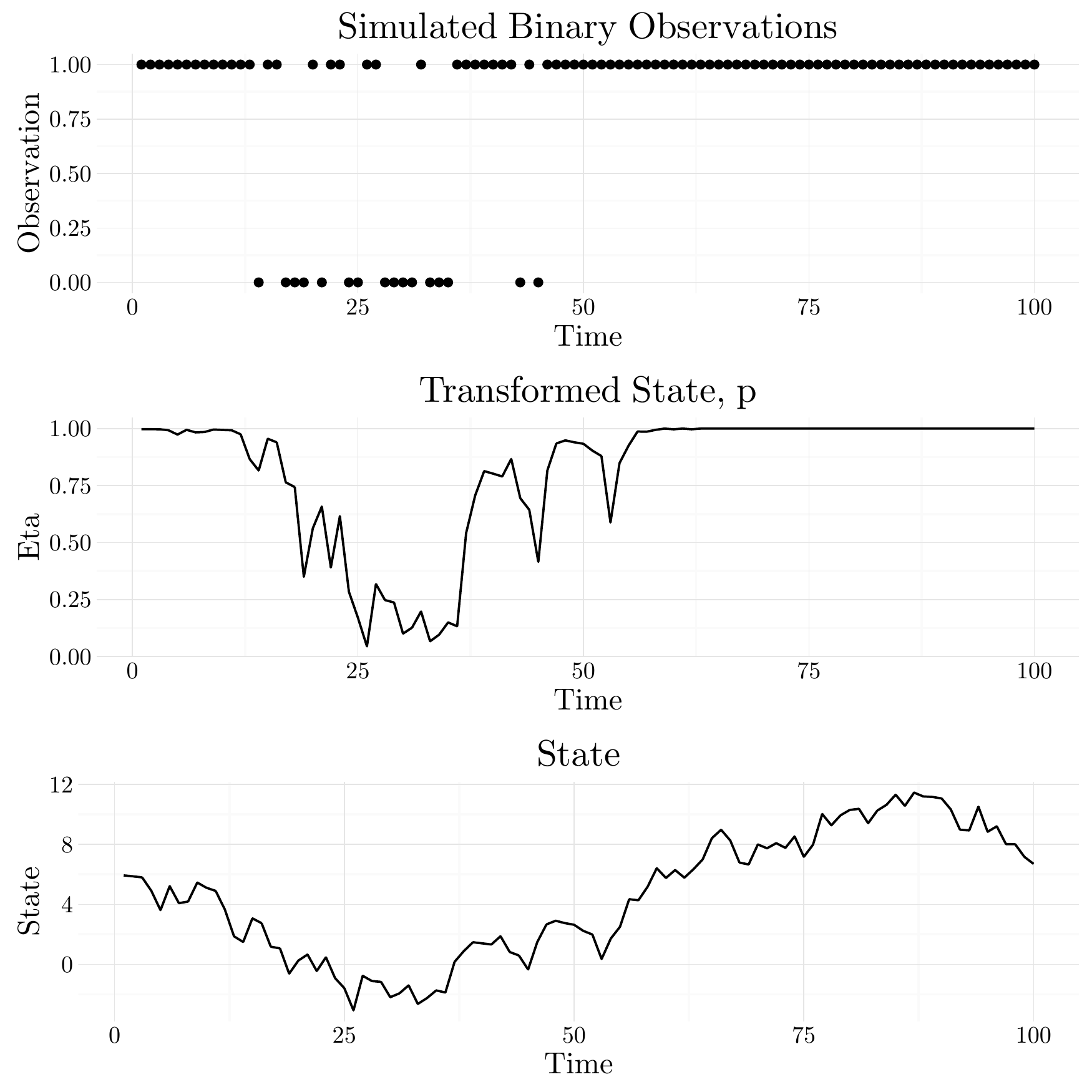}
\caption{Simulation of Bernoulli output, where the latent state $x(t)$ is a random walk with small positive drift. The binary outcomes are observed discretely on a regular integer lattice.\label{fig:bern}}
\end{figure}

The latent state, $x(t)$ is one-dimensional. Figure~\ref{fig:bern} shows a simulation from the Bernoulli POMP model on a regular integer lattice.

\subsubsection{Modelling Count Data: The Poisson Distribution}

The Poisson is one distribution used for modelling count data. The Poisson distribution can be represented as an exponential family model; let $S(Y(t_i)) = Y(t_i)$, $x(t_i) = \log\lambda(t_i)$, $a(x(t_i)) = \exp \{x(t_i)\}$, $b(Y(t_i), V) = \frac{1}{Y(t_i)!}$ and the scale parameter $V = 1$. The natural parameter, $x(t_i)$, is unobserved and varies in continuous time according to a Markov Process. The realm of $Y(t_i)$ is $\mathcal{Y} = \mathbb{Z}^+$, the positive integers including zero. The full specification is given as:

\begin{align*}
\pi(Y(t_i)|\lambda(t_i)) &= \textnormal{Poisson}(\lambda(t_i)), \\
\lambda(t_i)|x(t_i) &= \exp \{x(t_i)\}, \\
\textrm{d}X(t_i) &= \mu \textrm{d}t + \sigma \textrm{d}W(t_i).
\label{eqn:standardPoiss}
\end{align*}

\begin{figure}
\centering
\includegraphics[width=0.5\textwidth]{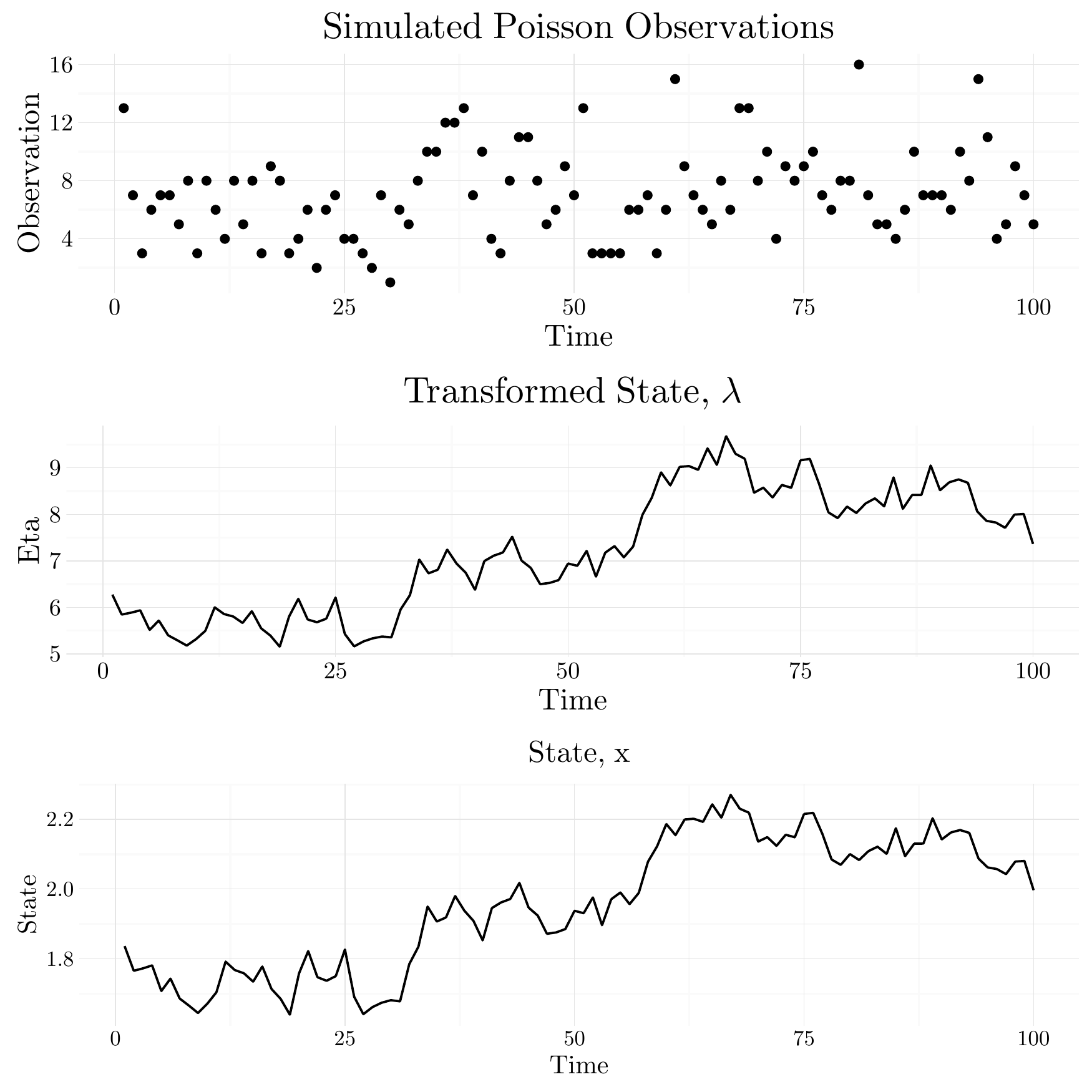}
\caption{(Top) Simulated Poisson data observed discretely at integer time points, (Middle) transformed unobserved state, $\lambda(t_i)$ (Bottom) State, $x(t_i)$, evolving according to generalised Brownian motion. \label{fig:pois}}
\end{figure}

The rate of the Poisson distribution, $\lambda(t_i)$, changes with time, and is constant between time steps. For instance, if count observations are aggregated in non-overlapping one hour windows, the rate is considered constant within each hour. A more realistic model of time-to-event data is presented later in Section~\ref{sec:lgcp}. Figure~\ref{fig:pois} shows simulation from a Poisson POMP Model (top) with a generalised Brownian Motion latent state (bottom) and the time varying rate of the Poisson distribution (middle).

\subsection{Modelling Seasonal Data}
\label{sec:seas}

Many natural phenomena feature predictable periodic changes in the rate of their process. For instance, when monitoring urban traffic, traffic flow is higher around 9am and 5pm as commuters make their way to and from the city. In order to model a periodic process, a time-dependent linear transformation vector of length $n$ is used:

\begin{equation*}
  F_{t_i} = \begin{pmatrix}
  	\cos(\omega t_i) \\
    \sin(\omega t_i) \\
    \vdots \\
    \cos(\omega h t_i) \\
    \sin(\omega h t_i)
    \end{pmatrix},
\end{equation*}
\noindent
where $\omega = \frac{2\pi}{T}$ is the frequency, $T$ represents the period of the seasonality and $h$ represents the number of harmonics required. The phase and amplitude of the waves are controlled by the value of the latent state, if the $h^{th}$ harmonic at time $t_i$ is given by:

\begin{equation*}
S_h = \begin{pmatrix}
\cos(\omega h t_i) \\
\sin(\omega h t_i)
\end{pmatrix} \cdot \begin{pmatrix}
x_1(t_i) \\
x_2(t_i)
\end{pmatrix},
\end{equation*}

\noindent
the phase of the wave is $\varphi = \arctan(-x_2(t_i)/x_1(t_i))$ and the amplitude is, $A = \sqrt{x_2(t_i)^2 + x_1(t_i)^2}$. This model has Normally distributed observations, and is given by:

\begin{align*}
\pi(Y(t_i)|\eta(t_i), V) &= \mathcal{N}(\eta(t_i), V), \\
\eta(t_i)|x(t_i) &= F_{t_i}^T x(t_i), \\
\textrm{d}\textbf{X}(t_i) &= \boldsymbol{\mu} \textrm{d}t + \Sigma \textrm{d}W(t_i).
\label{eqn:seasonalObservations}
\end{align*}

\begin{figure}
  \centering
  \includegraphics[width=0.5\textwidth]{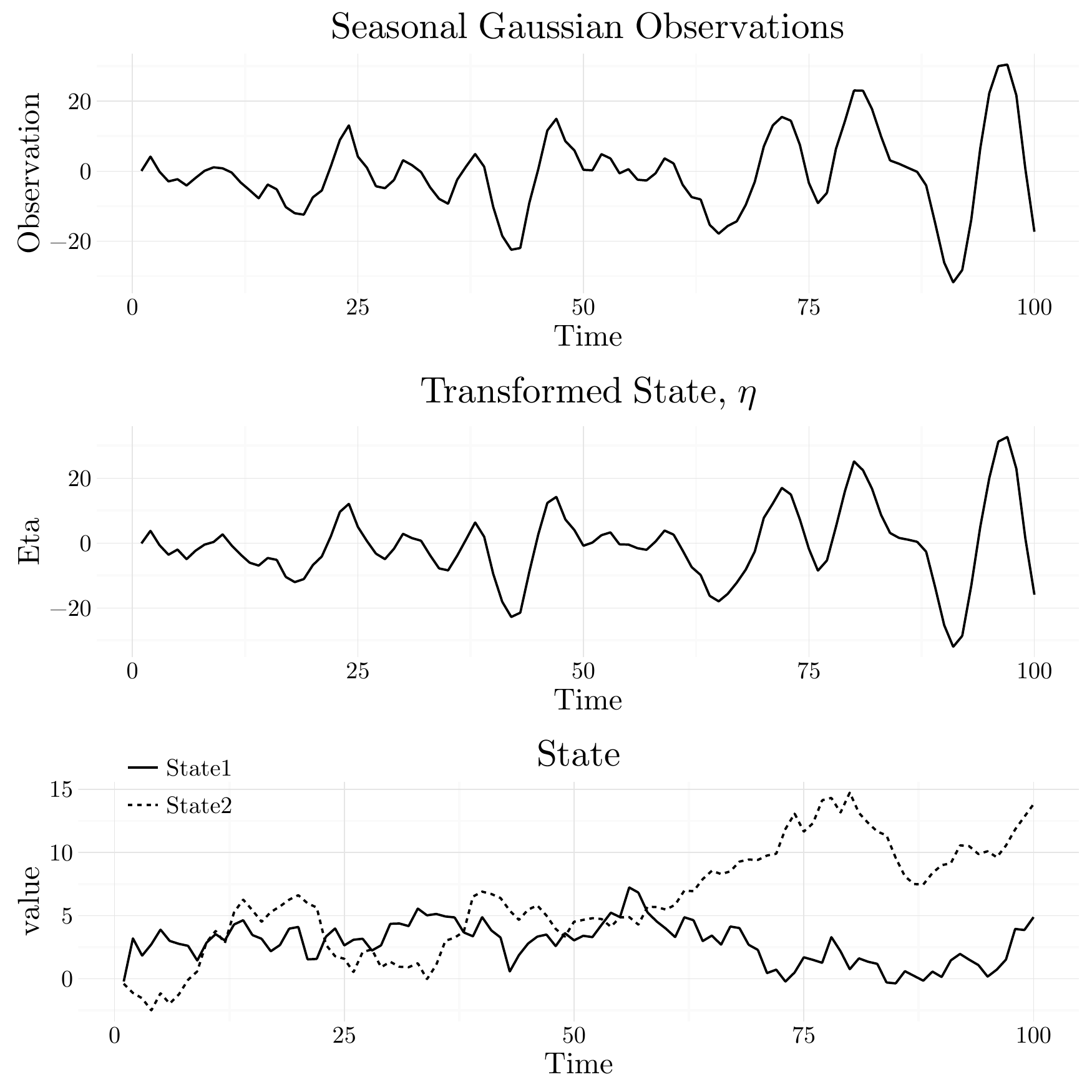}
  \caption{(Top) Simulated values from the Normal seasonal model, observed discretely at integer time points, (Middle) Transformed State, mean of the Normal Observations, (Bottom) Generalised Brownian Motion latent state \label{fig:brownSeas}}
\end{figure}

The latent state of the seasonal model is $2h$-dimensional. Figure~\ref{fig:brownSeas} shows a simulation from the seasonal model.

\subsection{Time to Event Data: The Log-Gaussian Cox-Process}
\label{sec:lgcp}

The Log-Gaussian Cox-Process (LGCP) is an inhomogeneous Poisson process whose rate is driven by a log-Gaussian process. In an inhomogeneous Poisson process the hazard, $\lambda(t)$, varies in time according to a log-Gaussian process. The total number of events in the interval $(0,T]$ is distributed as:

\begin{equation*}
N(t) \sim \text{Poisson}\left(\int_0^t \lambda(s) ds\right).
\end{equation*}

Denote the cumulative hazard as $\Lambda(t) = \int_0^t \lambda(s) ds$. The cumulative distribution function of the Log-Gaussian Cox-Process is $F(t) = 1 - \exp\{-\Lambda(t)\}.$ The density is then, 

\begin{equation}
\label{eqn:density}
f(t) = \frac{d}{dt}F(t) = \lambda(t)\exp\{-\Lambda(t)\}.
\end{equation}

\begin{figure}
\centering
\includegraphics[width=0.5\textwidth]{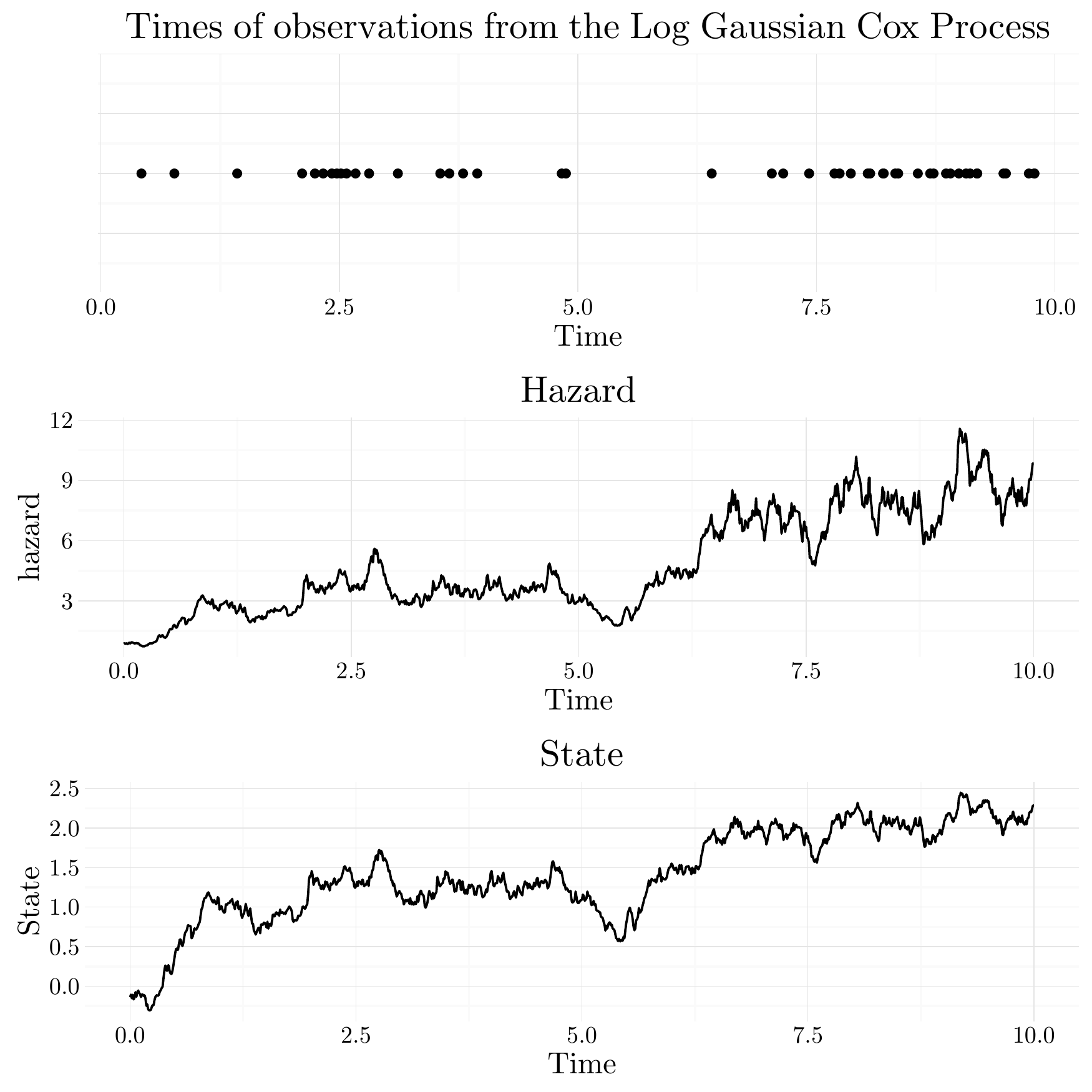}
\caption{(Top) Simulated event times from the Log-Gaussian Cox-Process with $\mu = 0.1$ and $\sigma = 0.5$, (Middle) The unobserved hazard, (Bottom) The unobserved state evolving according to generalised Brownian motion with positive drift \label{fig:lgcp}}
\end{figure}

In the Log-Gaussian Cox-Process, the hazard, $\lambda(t)$ is log-Gaussian. The general form of the LGCP POMP model is given by,

\begin{align*}
t|\lambda(t_i) &\sim \pi(t|\lambda(t_i), \Lambda(t_i)), \\
\begin{pmatrix}\lambda(t_i) \\ \textrm{d}\Lambda(t_i) \end{pmatrix} &= \begin{pmatrix} \exp\{x(t_i)\} \\ \lambda(t_i) \textrm{d}t \end{pmatrix}, \\
X(t_i)|x(t_{i-1}) &\sim p(x(t_i)|x(t_{i-1}), \theta), \quad x(t_0) \sim p_\theta(x(t_0))
\end{align*}

\noindent
where $p(x(t_i)|x(t_{i-1}), \theta)$ is the transition kernel of the Ornstein-Uhlenbeck process or other Gaussian Markov process. Figure~\ref{fig:lgcp} shows a simulation from the Log-Gaussian Cox-Process, using an approximate simulation algorithm presented in Appendix~\ref{sec:simlgcp}.

\subsection{The Latent State: Diffusion Processes}
\label{sec:state}

The system state evolves according to a continuous time Markov process. Any Markov process can be used to represent the state; however in this paper, the focus is on It\^{o} diffusion processes. Diffusion processes are represented by stochastic differential equations (SDE), (see~\cite{oksendal2013stochastic} for a detailed treatment). An SDE is a differential equation with a stochastic element, used to represent the continuous-time evolution of a stochastic random variable. An SDE of the form,

\begin{equation}
  \textrm{d}\textbf{X}(t) = \mu(\textbf{X}(t)) \textrm{d}t + \sigma(\textbf{X}(t)) \textrm{d}W(t),
\end{equation}

\noindent
is used to govern the time evolution of the Markovian latent state. Generally, $X \in \mathbb{R}^n$ and $\mu: \mathbb{R}^n \rightarrow \mathbb{R}^n$ is referred to as the drift coefficient, $\sigma: \mathbb{R}^n \rightarrow \mathbb{R}^{n\times m}$ is the diffusion coefficient and $W(t)$ is a Wiener Process.

If an SDE doesn't have an analytic solution, it can be simulated approximately using the Euler-Maruyama method~\citep{kloeden1992higher}. The Euler-Maruyama approximation gives an approximate solution to a stochastic differential equation, as a Markov chain. Start with a general SDE for a diffusion process:

\begin{equation*}
\textrm{d}X_t = \mu(X_t)\textrm{d}t + \sigma(X_t)\textrm{d}W_t.
\end{equation*}

The interval of interest, $(0,T]$ is partitioned into $n$ even sized sub-intervals, $\Delta t = T/n$. Define a value for $x(0)$, the initial value of the Markov chain, then

\begin{equation*}
X_{n+1} = X_n + \mu(X_n)\Delta t + \sigma(X_n)\Delta W_n,
\end{equation*}

\noindent
where $W_n \sim \mathcal{N}(0, \Delta t)$, are independent and identically distributed Normal random variables with mean 0 and standard deviation $\Delta t$. The approximate transition for any diffusion process can then by written as,

\begin{equation}
X_n|X_{n-1} \sim \mathcal{N}(X_{n-1} + \mu(X_n)\Delta t, \sigma(X_n)^2 \Delta t),
\end{equation}

\noindent
for sufficiently small $\Delta t$. The transition to the next state only depends on the current value, hence the Euler-Maruyama approximation scheme for stochastic differential equations is a Markov Process. Analytic solutions of diffusion processes are also Markovian, necessarily.

\section{Composing Models}
\label{sec:composed}

In order to model more complex phenomena, such as the traffic example presented in section~\ref{sec:poissoneg}, it is convenient to compose simple model components to form a more complex model. If the traffic is considered as count data, the observation distribution can be Poisson. Since the traffic data also displays daily and weekly seasonal cycles, then the instantaneous hazard, $\lambda(t)$ must vary periodically with time. In order to account for the two periods of seasonality with a Poisson observation model, a seasonal-Poisson model is formed by the composition of a Poisson model and two seasonal models, one with a weekly period and the other with a daily period. 

Consider an arbitrary model, with its associated functions, distributions, parameters and latent state indexed by $j$.

\begin{align*}
	Y(t_i)|\eta^{(j)}(t_i) &\sim \pi_j(y(t_i) | \eta^{(j)}(t_i), \theta^{(j)}), \\
	\eta^{(j)}(t_i)|\textbf{x}^{(j)}(t_i) &= g_j( F^{(j)T}_{t_i} \textbf{x}^{(j)}(t_i)), \\
	\textbf{X}^{(j)}(t_i)|\textbf{x}^{(j)}(t_{i-1}) &\sim p_j(\textbf{x}^{(j)}(t_i)|\textbf{x}^{(j)}(t_{i-1}), \theta^{(j)}), \quad \textbf{x}^{(j)}(t_0) \sim p_j(\textbf{X}^{(j)}(t_0) | \theta^{(j)}). \numberthis \label{eqn:generalmodel}
\end{align*}

Now define the composition of model $\mathcal{M}_1$ and $\mathcal{M}_2$ as $\mathcal{M}_3 = \mathcal{M}_1 \star \mathcal{M}_2$. Firstly consider the composition of observation models, by convention the observation model will be that of the left-hand model and the observations model of the right-hand model will be discarded. As such, the non-linear linking-function must be that of the left-hand model, $g_1: \mathbb{R} \rightarrow \mathcal{D}$. The linking-function ensures the state is correctly transformed into the parameter space for the observation distribution.

In order to compose the latent state, the initial state vectors are concatenated:

\begin{equation*}
\textbf{X}^{(3)}(t_o) \sim \begin{pmatrix} p(\textbf{x}^{(1)}(t_0) | \theta^{(1)}) \\
p(\textbf{x}^{(2)}(t_0) | \theta^{(2)}) \end{pmatrix}.
\end{equation*}

The composed model's transition function for the state is given by: 

\begin{equation*}
p_3(\textbf{x}(t_i) | \textbf{x}(t_{i-1}), \theta) = \begin{pmatrix}
p_1(\textbf{x}^{(1)}(t_i) | \textbf{x}^{(1)}(t_{i-1}), \theta^{(1)}) \\
    p_2(\textbf{x}^{(2)}(t_i) | \textbf{x}^{(2)}(t_{i-1}), \theta^{(2)})
\end{pmatrix}.
\end{equation*}

In order to compose the linear deterministic transformation vectors, $F^{(j)}(t)$, $j = \{1,2\}$, the vectors are concatenated. The vector dot-product with the latent state is then computed, so the result of applying the concatenated vector to the state is the same as applying each model vector to their respective state and adding the results: $F^{(3)T}_t \textbf{x}(t) \equiv F^{(1)T}_t \textbf{x}^{(1)}(t) + F^{(2)T}_t \textbf{x}^{(2)}(t)$. The dot product, $\gamma(t) = F_t^T \textbf{x}(t)$ results in a one-dimensional state, $\gamma(t) \in \mathbb{R}$. The full composed model, $\mathcal{M}_3$ can then be expressed as follows:

\begin{align*}
	Y(t_i)|\eta(t_i) &\sim \pi_1(y(t_i) | \eta(t_i)), \\
	\eta(t_i)|\textbf{x}(t_i) &= g_1(F^{(3)T}_{t_i} \textbf{x}(t_i)) \\
\textbf{X}(t_i) | \textbf{x}(t_{i-1})  &\sim \begin{pmatrix}
	p_1(\textbf{x}^{(1)}(t_i) | \textbf{x}^{(1)}(t_{i-1}), \theta^{(1)}) \\
    p_2(\textbf{x}^{(2)}(t_i) | \textbf{x}^{(2)}(t_{i-1}), \theta^{(2)})
\end{pmatrix}, \\ 
\textbf{X}(t_0) &\sim \begin{pmatrix}
p(\textbf{x}^{(1)}(t_0) | \theta^{(1)}) \\
p(\textbf{x}^{(2)}(t_0) | \theta^{(2)})
\end{pmatrix}.
\end{align*}

\begin{figure}
\tikzstyle{line} = [draw, -latex']
\tikzstyle{state} = [draw, circle, text width=1.5cm, align=center]
\tikzstyle{detstate} = [draw, double, circle, text width=1.5cm, align=center]

\begin{center} 
\begin{tikzpicture}[node distance = 2cm, auto]
    \node (1) {};
    \node [state, right of=1] (2) {$\textbf{x}^{(2)}(t_{i-1})$};
    \node [state, right of=2, node distance=3cm] (3) {$\textbf{x}^{(2)}(t_i)$};
    \node [state, right of=3, node distance=3cm] (4) {$\textbf{x}^{(2)}(t_{i+1})$};
    \node [right of=4, align=center] (5) {};
    \node [detstate, below of=2,node distance=3cm] (6) {$\eta(t_{i-1})$};
    \node [detstate, below of=3,node distance=3cm] (7) {$\eta(t_i)$};
    \node [detstate, below of=4, node distance=3cm] (8) {$\eta(t_{i+1})$};
    \node [state, below of=6, node distance=3cm] (9) {$y(t_{i-1})$};
    \node [state, below of=7, node distance=3cm] (10) {$y(t_i)$};
    \node [state, below of=8, node distance=3cm] (11) {$y(t_{i+1})$};
    \node [state, above of=2] (12) {$\textbf{x}^{(1)}(t_{i-1})$};
    \node [state, right of=12, node distance=3cm] (13) {$\textbf{x}^{(1)}(t_i)$};
    \node [state, right of=13, node distance=3cm] (14) {$\textbf{x}^{(1)}(t_{i+1})$};
    \node [right of=14] (15) {};
    \node [left of=12] (16) {};
    
    \draw [->] (12) to[out=225, in=135] (6);
    \draw [->] (13) to[out=225, in=135] (7);
    \draw [->] (14) to[out=225, in=135] (8);
    \path [line] (2) -- (3);
    \path [line] (3) -- (4);
    \path [line] (1) -- (2);
    \path [line] (4) -- (5);
    \path [line] (2) -- (6);
    \path [line] (3) -- (7);
    \path [line] (4) -- (8);
    \path [line] (6) -- (9);
    \path [line] (7) -- (10);
    \path [line] (8) -- (11);
    \path [line] (12) -- (13);
    \path [line] (13) -- (14);
    \path [line] (14) -- (15);
    \path [line] (16) -- (12);
    
    \node [above right=-0.4cm and 0.5cm of 2, node distance=1.5cm, fill=white] {$p_2$};
    \node [above right=-0.4cm and 0.5cm of 3, node distance=1.5cm, fill=white] {$p_2$};
    \node [above right=-0.4cm and 0.5cm of 12, node distance=1.5cm, fill=white] {$p_1$};
    \node [above right=-0.4cm and 0.5cm of 13, node distance=1.5cm, fill=white] {$p_1$};
    \node [rectangle, below=0.3cm of 2, fill=white] {$g_1(F^{(3)T}_{t_{i-1}} \textbf{x}^{(3)}(t_{i-1}))$};
    \node [rectangle, below=0.3cm of 3, fill=white] {$g_1(F^{(3)T}_{t_i} \textbf{x}^{(3)}(t_i))$};
    \node [rectangle, below=0.3cm of 4, fill=white] {$g_1(F^{(3)T}_{t_{i+1}} \textbf{x}^{(3)}_{t_{i+1}})$};
    \node [rectangle, below of=6, node distance=1.5cm, fill=white] {$\pi_1(y(t_{i-1})|\eta(t_{i-1}))$};
    \node [rectangle, below of=7, node distance=1.5cm, fill=white] {$\pi_1(y(t_{i})|\eta(t_{i}))$};
    \node [rectangle, below of=8, node distance=1.5cm, fill=white] {$\pi_1(y(t_{i+1})|\eta(t_{i+1}))$};
\end{tikzpicture}
\end{center}
\caption{A directed acyclic graph representing the composition of two models, the latent state is represented as separate, and advancing according to each model components transition kernel $p_i$ \label{fig:pompComposed}}
\end{figure}
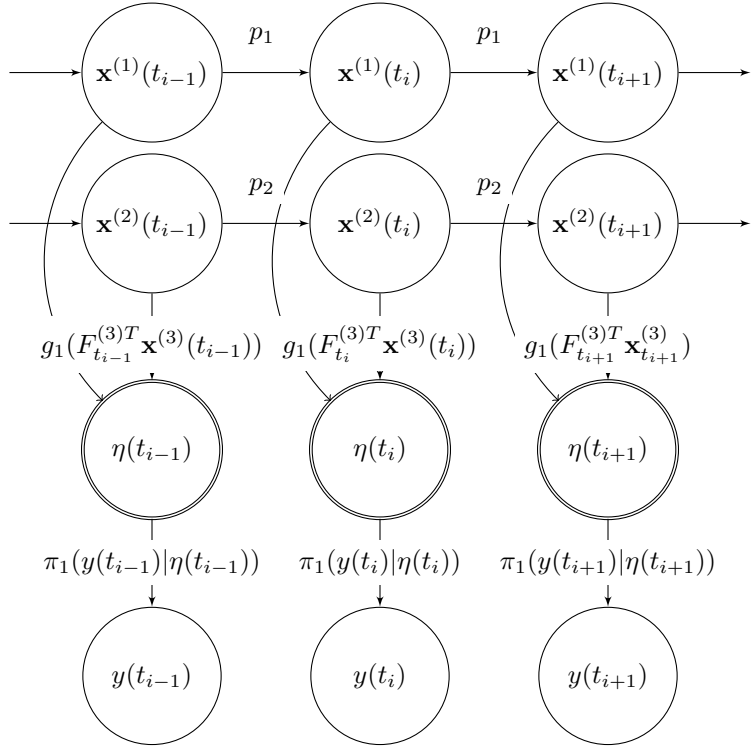

\noindent
Note that models form a semi-group under this composition operator, however the operator is not commutative. Figure~\ref{fig:pompComposed} shows a directed acyclic graph of a composed model. Further details on composing models in Scala are presented in appendix~\ref{sec:computingComposed}.

\subsection{Example: A Seasonal-Poisson Model}

To illustrate model composition, consider a Poisson model with a time dependent rate parameter, $\lambda(t)$ which varies seasonally. This model is the composition of two models, a single Poisson model, $\mathcal{M}_1$, and a single seasonal model, $\mathcal{M}_2$, to make $\mathcal{M}_3 = \mathcal{M}_1 \star \mathcal{M}_2$. This model could represent the flow of traffic through a city, as in the example in Section~\ref{sec:poissoneg}.

The Poisson model has a 1-dimensional state, which varies according to generalised Brownian motion. The linking function is the exponential function, $g(x) = \exp(x)$, and the linear transformation vector is the identity, $F^{(1)}_{t_i} = 1$. The Poisson model is given by:

\begin{align*}
N(t_i) &\sim \textnormal{Poisson}(\lambda(t_i)), \\
\lambda(t_i) | x(t_i) &= \exp\{ x(t_i) \}, \\
\textrm{d}X(t_i) &= \mu \textrm{d}t + \sigma \textrm{d}W(t_i).
\end{align*}

The latent state of the seasonal model has a dimension of $2h$, where $h$ represents the number of harmonics. Generalised Brownian motion is used to represent the time evolution of the $2h$-dimensional latent state, hence the drift coefficient $\boldsymbol{\mu}$ is $2h \times 1$ vector and the diffusion coefficient, $\Sigma$, is a $2h \times 2h$ diagonal matrix. The standard seasonal model has the Normal distribution as the observation distribution, and the linking function is the identity function:

\begin{align*}
N(t_i)|\lambda(t_i) &\sim \mathcal{N}(\eta(t_i), \sigma^2), \\
\eta(t_i)|\textbf{x}(t_i) &= F_{t_i}^{(2)T} \textbf{x}(t_i), \\
\textrm{d}\textbf{X}(t_i) &= \boldsymbol{\mu} \textrm{d}t + \Sigma \textrm{d}W(t_i)
\end{align*}

The vector $F^{(2)}_{t_i}$ is a $2h \times 1$ vector of fourier components:

\begin{equation}
F^{(2)}_{t_i} = \begin{pmatrix}
\cos \omega t \\
\sin \omega t \\
\cos 2\omega t \\
\sin 2\omega t \\
\vdots \\
\cos h\omega t \\
\sin h\omega t \\
\end{pmatrix},
\end{equation}

\noindent
$\omega = 2\pi/T$ is the frequency and $T$ is the period of the seasonality. In order to model a daily seasonality, $T = 24$, if the observation timestamps are in hour units.

In the composed model, the left-hand observation distribution is chosen, in this case the Poisson distribution. This necessitates the left-hand linking function ($g(x) = \exp(x)$) since the rate parameter of the Poisson distribution is greater than zero, $\lambda(t) > 0$. The full composed model can be expressed as:

\begin{align*}
N(t_i) | \lambda(t_i) &\sim \textnormal{Poisson}(\lambda(t_i)), \\
\lambda(t_i) | \textbf{x}(t) &= \exp\{F_{t_i}^{(3)T} \textbf{x}(t_i)\}, \\
\textrm{d}\textbf{X}(t_i) &= \boldsymbol{\mu}\textrm{d}t + \begin{pmatrix}
\sigma & 0 \\
0 & \Sigma
\end{pmatrix}\textrm{d}W(t_i).
\end{align*}

The system state is $n = 2h + 1$ dimensional, hence the drift coefficient $\boldsymbol{\mu}$ is $n \times 1$ vector and the diffusion coefficient is an $n \times n$ diagonal matrix. The linear transformation vector $F^{(3)}_{t_i}$ is a time-dependent, $n \times 1$ vector, which is the concatenation of the two $F^{(j)}_{t_i}, j = 1,2$, vectors:

\begin{equation}
\label{eqn:ft}
F^{(3)}_{t_i} = \begin{pmatrix}
1 \\
\cos \omega t \\
\sin \omega t \\
\vdots \\
\cos h\omega t \\
\sin h\omega t \\
\end{pmatrix}.
\end{equation}

\subsubsection{Example Composition in Scala}

Firstly, define a Poisson model with a \lstinline{stepFunction} representing the solution to a Markov process and the models associated parameters.

\begin{lstlisting}
val poissonMod: UnparamMod = PoissonModel(stepFunction)
val poissonParam: Parameters = LeafParameter(InitialStateParameters, None, MarkovParameters)
\end{lstlisting}

The \lstinline{LeafParameter} contains the parameters for the initial state and Markov transition kernel, the None in the middle corresponds to a scale parameter required by some observation distributions (however not required for the Poisson model). Now define a single seasonal model:

\begin{lstlisting}
val seasonalMod: UnparamMod = SeasonalModel(period = 24, harmonics = 3, stepFunction)
val seasonalParam: Parameters = LeafParameter(InitialStateParameters, None, MarkovParameters)
\end{lstlisting}

The seasonal model has a Gaussian observation model, in order to use the seasonal model by itself the measurement noise for the Gaussian observation model must be specified in the additional parameters (currently set as None). Now, composing the parameters and model:

\begin{lstlisting}
val combinedParams: Parameters = poissonParam |+| seasonalParam
val combinedModel: UnparamModel = poissonMod |+| seasonalMod
val parameterisedModel: Model = combinedModel(combinedParams)
\end{lstlisting}

The \lstinline{parameterisedModel} can be simulated from, it can be fit to observed data using the PMMH algorithm and particle filter and can used for online filtering as described in Section~\ref{sec:SI}.

\section{Statistical Inference}
\label{sec:SI}

Consider a POMP model of the form, observed at discrete times, $T = \{t_i | i = 1,\dots,M\}$:

\begin{align*}
	Y(t_i)|\eta(t_i) &\sim \pi(y(t_i) | \eta(t_i), \theta), \\
  	\eta(t_i)|\textbf{x}(t_i) &= g (F^T_{t_i} \textbf{x}(t_i)), \\
	\textbf{X}(t_i)|\textbf{x}(t_{i-1}) &\sim p(\textbf{x}(t_i)|\textbf{x}(t_{i-1}), \theta), \quad \textbf{x}(t_0) \sim p(\textbf{x}(t_0)|\theta).
\end{align*}

The joint distribution of all the random variables in the model can be written as:

\begin{align*}
p(\textbf{x}(t_{0:M}), y(t_{1:M}), \theta) &= p(\textbf{x}(t_{0:M}), \theta)\pi(y(t_{1:M}) | g(F^T_{t_{0:M}} \textbf{x}(t_{0:M})), \theta), \\
&= p(\theta) p(\textbf{x}(t_0)|\theta) \Big[\prod_{i=1}^M p(\textbf{x}(t_{i})|\textbf{x}(t_{i-1}), \theta) \pi(y(t_i) | \eta(t_i), \theta)\Big].
\end{align*}

Since the latent state is a Markov process, the joint distribution can be written as the product of the distribution over the parameters, $p(\theta)$, the initial distribution of the latent state $p(\textbf{x}(t_0)|\theta)$, the transition density, $p(\textbf{x}(t_{i})|\textbf{x}(t_{i-1}))$ and the likelihood $\pi(y(t_i) | \eta(t_i), \theta)$. In general this likelihood is analytically intractable, the bootstrap particle filter introduced in Section~\ref{sec:bootstrap}, can be used to calculate the pseudo-marginal likelihood and latent state of a POMP model, given observed data and parameters. Particle Marginal Metropolis-Hastings, presented in Section~\ref{sec:pmmh}, is used to determine the full joint-posterior distribution, $p(\textbf{x}(t_{0:M}), \theta | y(t_{0:M}))$, utilising the pseudo-marginal likelihood estimated using the bootstrap particle filter.

\subsection{State Estimation: The Bootstrap Particle Filter}
\label{sec:bootstrap}

The bootstrap particle filter \citep{Gordon1993} is a simple filter able to estimate the latent state of a POMP model. The hidden Markov process has an associated transition kernel $\textbf{X}(t_i)|\textbf{x}(t_{i-1}) \sim p(\textbf{x}(t_i)|\textbf{x}(t_{i-1}))$, from which realisations can be sampled. The transition kernel can either be an exact solution to an SDE or an SDE simulated on a fine grid using the Euler-Maruyama Method. The process is observed through an observation model $Y(t_i)|\eta(t_i) \sim \pi(y(t_i) | \eta(t_i))$.

The bootstrap particle filter is used to calculate the unobserved system state by approximating the filtering distribution, $p(\textbf{x}(t_i)| y(t_{0:i}))$. The algorithm to determine an empirical approximation of the unobserved system state at each observation time is presented below:

\begin{enumerate}
\item Initialise: Simulate $N$ particles from the initial distribution of the latent state, $\textbf{x}(t_0)^{(k)} \sim p(\textbf{x}(t_0))$, and set $i = 1$
\item Advance the particle-cloud until the time of the next observation, $\textbf{x}(t_i)^{(k)} \sim p(\textbf{x}(t_i)^{(k)}|\textbf{x}(t_{i-1})^{(k)}), k = 1,2,\dots,N$
\item Transform the each particle appropriately for the given observation model, $\eta(t_i)^{(k)} = g(F^T_{t_i} \textbf{x}(t_i)^{(k)})$
\item Calculate the weight of each particle, by calculating the likelihood of the observation given each particle: $w^*(t_i)^{(k)} = \pi(y(t_i)|\eta(t_i)^{(k)})$
\item Normalise the weights, $w(t_i)^{(k)} = \frac{w^*(t_i)^{(k)}}{\sum_{j=1}^N w^*(t_i)^j}$. The particles and associated normalised weights form a weighted sample from the filtering distribution $p(\textbf{x}(t_i)| y(t_{1:i}))$, $\{\textbf{x}(t_i)^{(k)}, w(t_i)^{(k)} | k = 1,\dots,N \}$
\item Resample the state, by sampling with replacement $N$ times from a Multinomial distribution with each category representing a particle and the probability of choosing a particle represented by the associated weights, $w(t_i)^{(k)}$. This gives an approximate random sample from $p(\textbf{x}(t_i)| y(t_{0:i}))$
\item Return the random samples from the filtering distribution $p(\textbf{x}(t_i)| y(t_{0:i})).$ If $i < M$ set $i = i + 1$ and go to step 2, else stop.
\end{enumerate} 

The average of the un-normalised weights at each time point gives an estimate of the marginal likelihood of the current data point given the data observed so far:

\begin{equation*}
\hat{p}(y(t_i)|y(t_{1:i-1})) = \frac{1}{N}\sum_{j=1}^N w^*(t_i)^j.
\end{equation*}

The estimate of the likelihood of the full path is given by:

\begin{equation*}
\hat{p}(y(t_{1:M})) = \hat{p}(y(t_1))\prod_{i=2}^M \hat{p}(y(t_i)|y(t_{1:i-1})).
\end{equation*}

The estimated marginal likelihood is consistent, meaning that as the number of particles are increased, $N \rightarrow \infty$, then the estimate converges in probability to the true value. The estimate is also unbiased, meaning $\mathbb{E}(\hat{p}_\theta^*(y(t_{1:M}))) = p_\theta^*(y(t_{1:M}))$, see \cite{del2004feynman} for a proof. This marginal likelihood is used in the Particle Marginal Metropolis Hastings (PMMH) algorithm discussed in Section~\ref{sec:pmmh}.

In practice, the bootstrap particle filter can be easily parallelised, advancing the state and calculating the likelihood are naturally parallel. However, Multinomial resampling (as described above) requires communication among multiple threads to sum the value of the weights. Other resampling schemes, such as stratified resampling and systematic resampling are more amenable to parallelisation. An overview of resampling schemes used in the particle filter are considered by~\cite{Murray2016}.

\subsubsection{Implementation (Fold and Scan)}
\label{sec:bootImpl}

The models and algorithms described in this paper have been implemented in Scala, a functional, object-oriented language which runs on the Java Virtual Machine (JVM). This means the code can be deployed without change across different operating systems and on servers in the cloud which have the JVM installed. Observations arrive as a stream, a stream can be thought of as a lazy list. A list can be represented recursively as a pair, with the first element a single value at the head of the list and the second element another list (called the tail), this is called a cons list. A lazy list, or stream, is also a pair, with the first element being a computed value and the second element a function with no parameters (in Scala, a \lstinline{lazy val}). The stream can be forced to evaluate its contents, for instance when performing a side effect such as printing to the screen or writing to a file. In practice, when considering live streaming data, the function in the second element of the pair might be one which polls a web service for the next observation. The approach taken in this paper is similar to \citeauthor{Beckman2016}'s series of papers on Kalman folding, implementing the Kalman filter on streams in the Wolfram language~\citep{Beckman2016}.

\lstinline{foldLeft} is a function which operates on a recursive datatype, such as list or stream. The function signature of \lstinline{foldLeft} is given by:

\begin{lstlisting}
def foldLeft[A, B](l: Foldable[A], z: B)(f: (B, A) => B): B
\end{lstlisting}

\noindent
\lstinline{foldLeft} is used to apply a function to elements of a \lstinline{Foldable} (\lstinline{Stream} is a \lstinline{Foldable} data structure) pairwise in order to calculate an aggregated result. The function \lstinline{f} takes two arguments, the first of which is of type \lstinline{B} (a placeholder for any type) and the second is of type \lstinline{A}, the same as the element type of the Stream (whatever that may be). The first application of \lstinline{f}, is to the first element of the stream and \lstinline{z} (the so called zero of the fold). The result is then passed to the next application of \lstinline{f}, along with the second element of the stream.

For instance, consider adding values in a list using foldLeft:

\begin{lstlisting}
val l = List(1,2,3,4,5)
val sum2 = (a, b) => a + b
foldLeft(l, 0)(sum2)
\end{lstlisting}

The first application of \lstinline{sum2} will add the zero element, 0, to the first element in the list (from the left): \lstinline{sum2(0, 1) = 0 + 1 = 1}. Then the next element of the list and the previously evaluation of \lstinline{sum2} is passed to the next evaluation of \lstinline{sum2}: \lstinline{sum2(1, 2) = 1 + 2 = 3}. This is repeated until the list is exhausted and the value returned is 15. In this way, the list is summed without mutating state in the program. Also note that if \lstinline{A} = \lstinline{B} and the function \lstinline{f: (A, A) => A} is associative, the fold can be computed in parallel via a binary tree reduction.

In the application of the bootstrap particle filter there is an internal state which propagates as the stream of data arrives. The internal state includes the particle cloud, the time of the most recent observation and the log-likelihood. The function \lstinline{f} in \lstinline{foldLeft} can be implemented as a single step of the particle filter. An illustrative implementation of \lstinline{f}, called \lstinline{filterStep} is implemented in the code block~\ref{lst:particleFilter}. Firstly a \lstinline{trait} is defined containing abstract implementations of three important functions in the particle filter, which will be implemented when applying the particle filter in practice. \lstinline{stepFunction} and \lstinline{dataLikelihood} are implemented in each model and can be specified in a concrete class for a specific particle filter implementation. The \lstinline{resample} function is not model specific, and Multinomial resampling is typically employed.

\begin{lstlisting}[caption={Illustrative implementation of a particle filter, for use in reduction functions \lstinline{foldLeft} and \lstinline{scanLeft}},label={lst:particleFilter}]
trait ParticleFilter {
    import math._
    import ParticleFilter._

    val resample: (Vector[State], Vector[LogLikelihood]) => Vector[State]
    val stepFunction: (State, TimeIncrement) => State
    val dataLikelihood: (State, Observation) => LogLikelihood
    val mean: Vector[LogLikelihood] => LogLikelihood = w => w.sum/w.length

    def filterStep(s: FilterState, y: Data): FilterState = {
        val dt = y.time - s.t0
  
        val x = resample(s.particles, s.weights)

        val x1 = s.particles map (stepFunction(_, dt))
        val w = x1 map (dataLikelihood(_, y.observation))

        val ll = s.ll + log(mean(w map (exp)))
 
        FilterState(x1, w, y.time, ll)
    }
}

object ParticleFilter {
    type Time = Double
    type Observation = Double
    type LogLikelihood = Double
    type State = Vector[Double]
    type TimeIncrement = Double

    case class FilterState(
        particles: Vector[State], 
        weights: Vector[LogLikelihood], 
        t0: Time, 
        ll: LogLikelihood)
    
    case class Data(time: Time, observation: Observation)
}
\end{lstlisting}

In the function \lstinline{filterStep}, firstly, the difference between subsequent observations is calculated, \lstinline{dt}. Then the states are resampled to get an approximate unweighted random sample from $p(\textbf{x}(t_i)|y(t_{1:i}))$. Each particle is advanced independently by applying the function \lstinline{stepFunction} to each particle using the higher-order function \lstinline{map} which has the following signature:

\begin{lstlisting}
def map[B](l: Vector[A])(f: (A) => B): Vector[B]
\end{lstlisting}

\lstinline{map} is used to apply a function to the inner type of the \lstinline{Vector}, in this case \lstinline{A}. Each particle weight is calculated independently using \lstinline{map} and \lstinline{dataLikelihood}. In practice, we compute the log-likelihood to avoid arithmetic underflow from multiplying many small values. The value of the log-likelihood, $\hat{p}_\theta(y(t_{1:M}))$, is updated by adding the log of the mean of the weights to the previous value.

In order to calculate the log-likelihood (ie a single value) of a path, the function \lstinline{foldLeft} can be employed:

\begin{lstlisting}[caption={Scala Code to calculate the log-likelihood of a set of observations},label={lst:calculateLL}]
val mod = // model here
val data: Stream[Data] = // stream of data here
val initState = FilterState(
    mod.x0.sample(1000), 
    Vector.fill(1000)(1.0/1000),
    0.0,
    0.0)
    
foldLeft(data, initState)(filterStep).ll

\end{lstlisting}

\noindent
the value \lstinline{initState} is implemented by sampling 1,000 times (equivalent to 1,000 particles) from the initial state distribution of the model and initialising the weights to be 1/1,000. The initial time \lstinline{t0} is taken to be 0.0 and the initial value of the log-likelihood is set at zero by convention. The log-likelihood is extracted by appending \lstinline{.ll} on the call to \lstinline{foldLeft}.

A closely related function to \lstinline{foldLeft} is \lstinline{scanLeft} which will return an aggregated stream of reduced values using a provided function, the signature is:

\begin{lstlisting}
def scanLeft[A, B](l: Stream[A], z: B)(f: (B, A) => B): Stream[B]
\end{lstlisting}

\noindent
In order to understand \lstinline{scanLeft}, consider the application of \lstinline{sum2} to a stream of natural numbers:

\begin{lstlisting}
val naturalNumbers = Stream.from(1)
scanLeft(naturalNumbers, 0)(sum2)
\end{lstlisting}
\noindent
then an infinite stream containing the cumulative sum will be returned, \lstinline{0, 1, 3, 6, ...}. The following code block runs a particle filter on a stream of data, where filterStep, initState and data are the same as supplied above in~\ref{lst:calculateLL}.

\begin{lstlisting}
scanLeft(data, initState)(filterStep)
\end{lstlisting}

\noindent
this code accumulates the particle cloud (states and associated weights) and the running log-likelihood into a stream, using the reduction function \lstinline{filterStep}.

\subsubsection{Filtering for the Log-Gaussian Cox-Process}
\label{sec:lgcpfiltering}

When considering observations from the Log-Gaussian Cox-Process (LGCP), the filtering needs to be performed slightly differently. The likelihood for the LGCP is given by:

\begin{equation*}
\pi(y(t) | \lambda(t), \Lambda(t)) = \lambda(t) \exp(-\Lambda(t)),
\end{equation*}

\noindent
notice that the likelihood depends on the instantaneous hazard $\lambda(t)$ and the cumulative hazard $\Lambda(t) = \int_0^t \lambda(s) ds$. In practice the log-likelihood is calculated to avoid arithmetic overflow and to simplify calculations, the log likelihood is given by, $\ell = \log(\lambda(t)) - \Lambda(t)$. The state must be augmented to include $\Lambda(t)$ in addition to $\lambda(t)$. In practice the value of $\Lambda(t)$ is often calculated approximately, using numerical integration.

\subsection{Parameter Estimation}
\label{sec:pmmh}
The Particle Marginal Metropolis-Hastings algorithm (PMMH)~\citep{Andrieu2010} is an offline Markov chain Monte Carlo (MCMC) algorithm which targets the full joint posterior $p(\textbf{x}(t_{0:M}), \theta | y(t_{1:M}))$ of a partially observed Markov process. Consider a POMP model given by,

\begin{align*}
Y(t_i)|\eta(t_i) &\sim \pi(y(t_i) | \eta(t_i), \theta), \\
\eta(t_i)|\textbf{x}(t_i) &= g(F^T_{t_i} \textbf{x}(t_i)), \\
\textbf{X}(t_i)|\textbf{x}(t_{i-1}) &\sim p(\textbf{x}(t_i) | \textbf{x}(t_{i-1}), \theta), \quad \textbf{x}(t_0) \sim p(\textbf{x}(t_0)|\theta),
\end{align*}

\noindent
where $\theta$ represents the parameters to be estimated using the PMMH algorithm. The parameters include the measurement noise in the observation distribution, $\pi(y(t_i) | \eta(t_i), \theta)$, the parameters of the Markov transition kernel for the system state, $p(\textbf{x}(t_i) | \textbf{x}(t_{i-1}), \theta)$ and the parameters of the initial state distribution $p(\textbf{x}(t_0)|\theta)$. The data, $y(t)$, is observed discretely. In order to simulate a Markov chain which targets the full posterior, $p(\theta, \textbf{x} | y)$, firstly a new set of parameters $\theta^*$ is proposed from a proposal distribution $q(\theta^*|\theta)$. Then the bootstrap particle filter (see Section~\ref{sec:bootstrap}), is run over all of the observed data up to time $t$ using the newly proposed $\theta^*$. The output of running the filter with the new set of parameters, $\theta^*$ is used to estimate the marginal likelihood, $\hat{p}_{\theta^*}(y) = \prod_{i=1}^n \hat{p}_{\theta^*}(y(t_i)|y(t_{i-1}))$ and, optionally, to sample a new proposed system state, $x^*$ from the conditional distribution $p(x^* | \theta^*, y)$. The pair $(\theta^*, x^*)$ are accepted with probability $\text{min}(1, A)$, where $A$ is given by:

\begin{equation}
\label{eqn:metropRatio}
A = \frac{p(\theta^*)\hat{p}_{\theta^*}(y) q(\theta^*|\theta)}{p(\theta)\hat{p}_{\theta}(y)q(\theta|\theta^*)},
\end{equation}

\noindent
the distribution $p(\theta)$, represents the prior distribution over the parameters. It is shown in \cite{Andrieu2010} that this algorithm has as its target the exact posterior $p(\theta|y)$, or optionally, $p(\theta, x | y)$.

The Metropolis-Hastings Kernel, can be simplified in the case of a symmetric proposal distribution. For a symmetric proposal distribution $q(\theta^*|\theta) = q(\theta|\theta^*)$. Commonly, the proposal is chosen to be a Normal distribution centered at the previously selected parameter, this is know as a random walk proposal, $q(\theta^*|\theta) = \mathcal{N}(\theta, \sigma)$, where $\sigma$ is a parameter controlling the step size of the random walk. If a flat prior distribution is chosen, then the ratio in Equation~\ref{eqn:metropRatio} can be simplified further to:

\begin{equation*}
A = \frac{\hat{p}_{\theta^*}(y)}{\hat{p}_{\theta}(y)}.
\end{equation*}

The full-joint posterior distribution is explored by performing many iterations of the PMMH algorithm, discarding burn-in iterations and possibly thinning the iterations to get less correlated samples from the posterior distribution.

\subsubsection{Implementation (MCMC as a stream)}

The Particle Marginal Metropolis-Hastings algorithm must be applied to a batch of data. Window functions, such as \lstinline{grouped}, can be applied to a stream of data to aggregate observations into a batch. \lstinline{grouped} accepts an integer, $n$, and groups each observation into another (finite) stream of size $n$.

The PMMH algorithm can then be applied to the aggregated group using \lstinline{map}. Iterations from the PMMH algorithm are naturally implemented as a stream. In the Scala standard library there is a method for producing infinite streams from an initial seed:

\begin{lstlisting}
def iterate[A](start: A)(f: A => A): Stream[A] 
\end{lstlisting}

\lstinline{iterate} applies the function \lstinline{f} to the starting value, then passes on the result to the next evaluation of \lstinline{f}. For example, to create an infinite stream of natural numbers:

\begin{lstlisting}
val naturalNumbers: Stream[Int] = iterate(1)(a => a + 1)
\end{lstlisting}

Iterations of an MCMC algorithm can be generated using \lstinline{iterate}, by starting with an initial value of the required state (at a minimum the likelihood and the initial set of parameters) and applying the Metropolis-Hastings update at each iteration. Inside of each application of \lstinline{f}, a new value of the parameters is proposed, the marginal likelihood is calculated using the new parameters (using the bootstrap particle filter) and the Metropolis-Hastings update is applied.

An illustrative example of a single step in the PMMH algorithm using the Metropolis Kernel is in code block~\ref{lst:metropHastings}. Three important functions are given abstract implementations in the \lstinline{MetropolisHastings} trait, \lstinline{proposal}, \lstinline{prior} and \lstinline{logLikelihood}. The \lstinline{proposal: Parameters => Rand[Parameters]} is a function representing the (symmetric) proposal distribution, \lstinline{Rand} is a representation of a distribution which can be sampled from by calling the method \lstinline{draw}. \lstinline{logLikelihood: Parameters => LogLikelihood} is a particle filter, with the observed data and number of particles fixed, which outputs an estimate of the log-likelihood for a given value of the parameters. \lstinline{prior: Parameters => LogLikelihood} represents the prior distribution over the parameters. These three functions will be implemented in a concrete class extending the \lstinline{MetropolisHastings} trait and correspond to specific implementation of the PMMH algorithm.

\begin{lstlisting}[caption={Illustrative implementation of the PMMH, with an implementation of \lstinline{stepMetrop} for use in \lstinline{Stream.iterate}},label={lst:metropHastings}]
trait MetropolisHastings {
    import MetropolisHastings._
    import math._
    import breeze.stats.distributions.{Rand, Uniform}

    val prior: Parameters => LogLikelihood
    val proposal: Parameters => Rand[Parameters]
    val logLikelihood: Parameters => LogLikelihood

    val stepMetrop: MetropState => MetropState = s => {
        val propParams = proposal(s.params).draw
        val propll = logLikelihood(propParams)
        val a = propll + prior(propParams) - s.ll - prior(s.params)

        if (log(Uniform(0, 1).draw) < a)
            MetropState(propll, propParams)
        else s
    }
}

object MetropolisHastings {
    type Parameters = Vector[Double]
    type LogLikelihood = Double

    case class MetropState(ll: LogLikelihood, params: Parameters) 
}
\end{lstlisting}

In order to generate a stream of iterations, use \lstinline{iterate}:

\begin{lstlisting}
val initState = MetropState(-1e99, initParams)
val iters = iterate(initState)(stepMetrop)
\end{lstlisting}

Where \lstinline{initParams} are drawn from the prior distribution and the initial value of the log-likelihood is chosen to be very small so the first iteration of the PMMH is accepted.

Built in stream operations can be used to discard burn-in iterations and thin the iterations to reduce auto-correlation between samples. The stream can be written to a file or database at each iteration, so the PMMH algorithm implemented as a stream uses constant memory as the chain size increases.

\section{Examples}
\label{sec:example}

In this section two datasets are considered, the first consists of traffic arrival time at various stations around Newcastle Upon Tyne. The second consists of urban temperature measurements recorded on a sensor network also deployed around Newcastle Upon Tyne in the Urban Observatory~\citep{uo2014Temperature}.

\subsection{UTMC Traffic Data: Seasonal-Poisson Model}
\label{sec:poissoneg}

\begin{figure}
\centering
\includegraphics[width=0.8\textwidth]{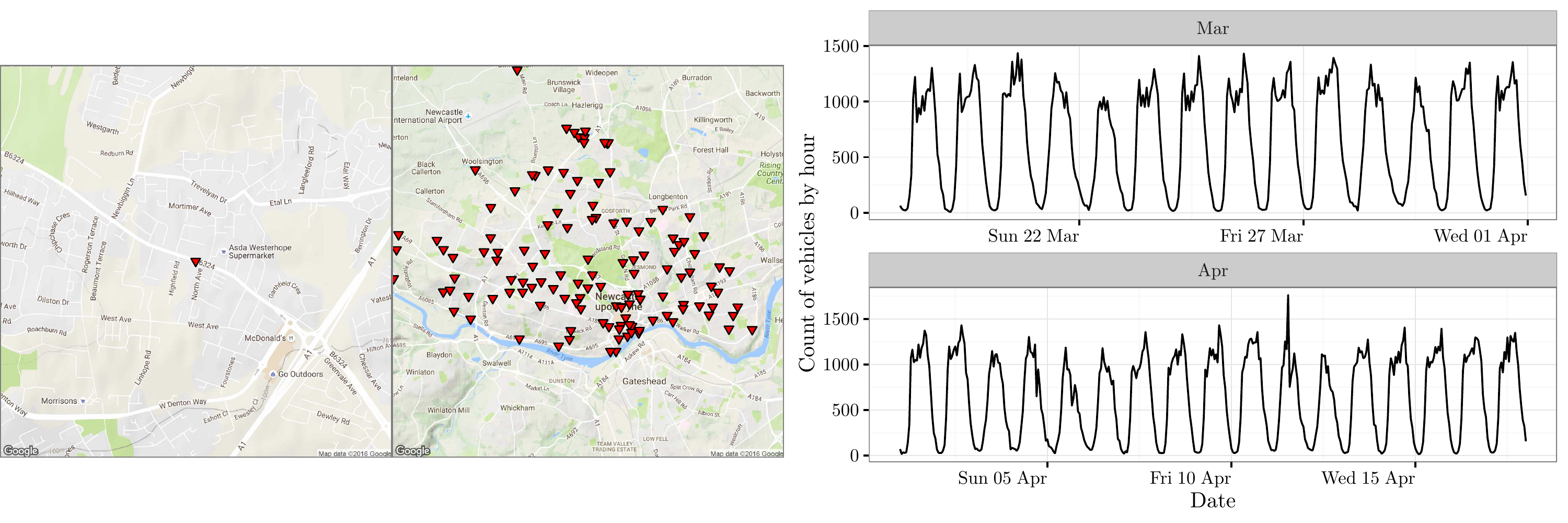}
\caption{(Left) sensor number 62, located on Stamfordham Road (Centre) All traffic sensors in Newcastle Upon Tyne (Right) Count of cars arriving per hour during part of March and April 2015 at station 62 on Stamfordham Road\label{fig:allsensors}}
\end{figure}

The Urban Traffic Management Control~\footnote{North East Combined Authority, Open Data Service. https://www.netraveldata.co.uk/}, operates a grid of sensors deployed across the five districts of Tyne and Wear with the aim to analyse traffic flow in order to prevent congestion. The sensor locations are plotted on the map in Figure~\ref{fig:allsensors}.

The raw data consists of a timestamp indicating the time the vehicles passed a monitoring station along with their direction, lane and speed. A useful way to summarise the data is by aggregating the count of cars per hour; counts of cars per hour at a station on Stamfordham Road are plotted on the right of Figure~\ref{fig:allsensors}. There is a strong daily seasonal component, with a slightly less pronounced weekly component. There is a distinct systematic drop in traffic on weekends, this coincides with fewer commuters on the road. Along with the daily and weekly seasonality, there are two peaks each day, coinciding with rush hour in the morning and the evening. Grouping the data this way naturally leads to a Poisson observation model.

In order to model this data, a composition of three models is used. The first model in the composition is Poisson:

\begin{align*}
N(t) &\sim \textnormal{Poisson}(\lambda(t)), \\
\lambda(t) | x(t) &= \exp\{x(t)\}, \\
\textrm{d}X(t) &= \mu\textrm{d}t + \sigma\textrm{d}W(t). \numberthis \label{eqn:poissonExampleModel}
\end{align*}

\noindent
The observation distribution is Poisson, hence the linking function is the exponential function, the linear-transformation vector is the identity, $F^{(1)}_t = 1$, the latent state is one-dimensional and evolves continuously according to generalised Brownian motion.

The second model in the composition is a seasonal model, representing daily seasonality:

\begin{align*}
Y(t) &\sim \mathcal{N}(\mu(t), V), \\
\mu(t) | \textbf{x}(t) &= F_t^{(2)T} \textbf{x}(t), \\
\textrm{d}X(t) &= \boldsymbol{\alpha} (\boldsymbol{\theta} - \textbf{X}(t))\textrm{d}t + \Sigma \textrm{d}W(t). \numberthis \label{eqn:seasonalExampleModel}
\end{align*}

\noindent
The linking function is the identity function, the observation distribution is Gaussian. The state is $2h_1$-dimensional, where $h_1$ is the chosen number of harmonics, and evolves according to a multivariate Ornstein-Uhlenbeck process. The linear-transformation vector is given by:

\begin{equation*}
F^{(2)}_t = \begin{pmatrix}
\cos \omega_1 t \\
\sin \omega_1 t \\
\cos 2 \omega_1 t \\
\sin 2 \omega_1 t \\
\cos 3\omega_1 t \\
\sin 3\omega_1 t 
\end{pmatrix},
\end{equation*}

\noindent
where $\omega_1 = 2\pi/T_1$ and $T_1 = 24$, since the data is aggregated in hourly intervals. There are three harmonics in the model.

The third model component in the composition is another seasonal model, representing weekly seasonality. This model component is the same as the model in equation~\ref{eqn:seasonalExampleModel}, but the period of the seasonality is $T_2 = 24 \times 7$, hence $\omega_2 = 2\pi/T_2$.

The model is then composed, with the observation distribution and linking-distribution taken from the left-hand (Poisson) model:

\begin{align*}
N(t_i) &\sim \textnormal{Poisson}(\lambda(t_i)) \\
\lambda(t_i) | \textbf{x}(t_i) &= \exp\{F_{t_i}^T \textbf{x}(t_i)\}  \\
\textbf{X}(t_i) | \textbf{x}(t_{i-1}) &= \begin{pmatrix}
p_1(x^{(1)}(t_i)| x^{(1)}(t_{i-1}), \theta) \\
p_2(\textbf{x}^{(2)}(t_i)| \textbf{x}^{(2)}(t_{i-1}), \theta) \\
p_3(\textbf{x}^{(3)}(t_i)| \textbf{x}^{(3)}(t_{i-1}), \theta)
\end{pmatrix}. \numberthis \label{eqn:model}
\end{align*}

The latent state from each model is concatenated, and each models state advances according to its own transition kernel. The vector, $F_t$ is the concatenation of the three linear-transformation vectors from Poisson, daily seasonal and weekly seasonal model:

\begin{equation}
\label{eqn:seasonality}
F_t = \begin{pmatrix}
1 \\ 
\cos \omega_1 t \\
\sin \omega_1 t \\
\cos 2 \omega_1 t \\
\sin 2 \omega_1 t \\
\cos 3 \omega_1 t \\
\sin 3 \omega_1 t \\
\cos \omega_2 t \\
\sin \omega_2 t \\
\cos 2 \omega_2 t \\
\sin 2 \omega_2 t \\
\cos 3 \omega_2 t \\
\sin 3 \omega_2 t
\end{pmatrix}.
\end{equation}

The parameter posterior distribution is determined by running the PMMH algorithm on a batch of 1,000 data points (each data point is the count of vehicles passing station 62 in an hour), consisting of approximately six weeks of data.

\begin{figure}
\centering
\includegraphics[width=0.5\textwidth]{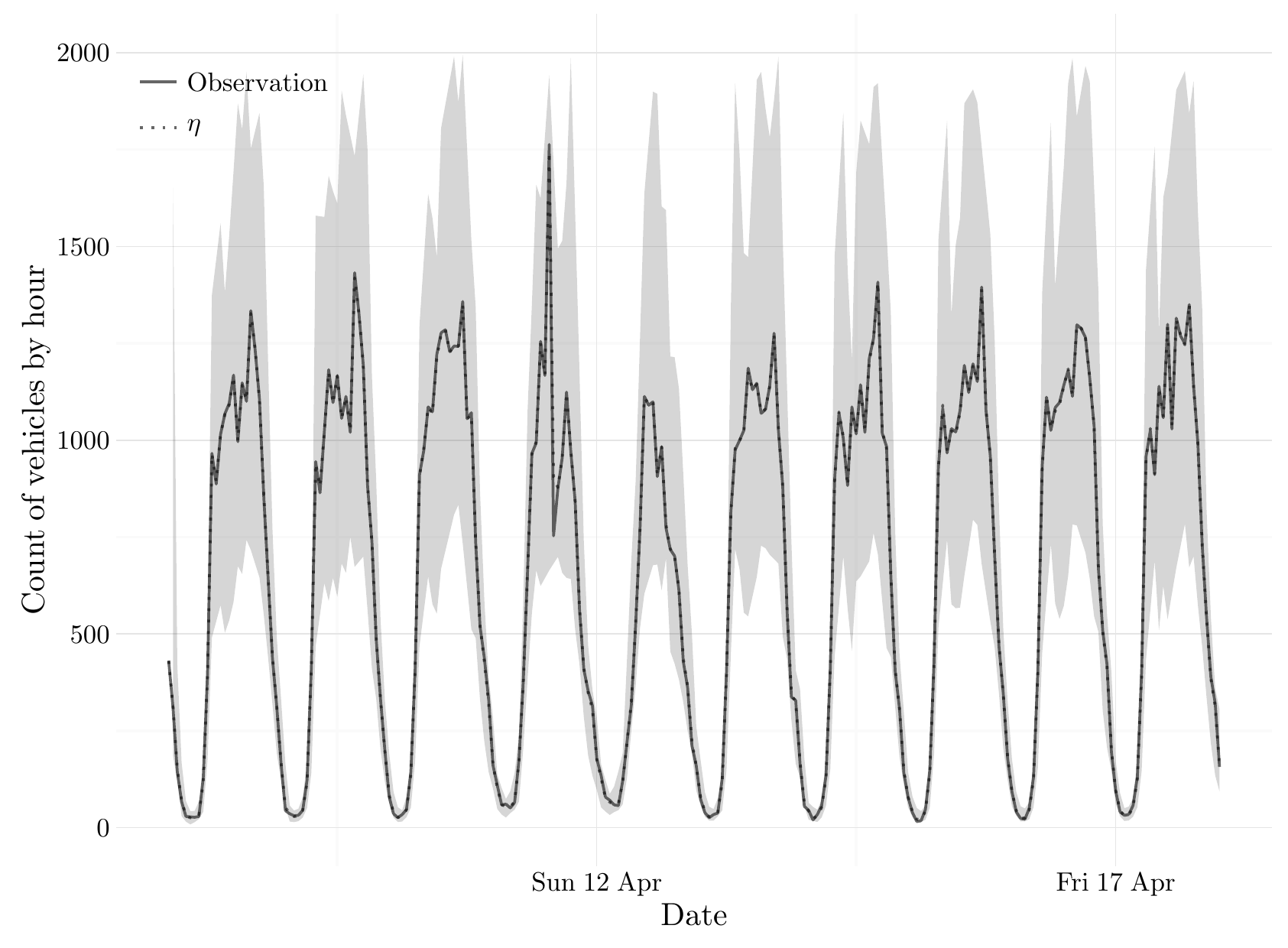}
\caption{The mean of the filtering distribution and associated 99\% credible intervals are plotted alongside the actual observations, the mean of the filtering distribution very closely aligns with the actual observations, which are assumed to be drawn from a Poisson distribution with this mean \label{fig:filteringDist}}
\end{figure}

The mean of the parameter posterior distributions can be used to run a particle filter on future observations to determine the filtering distribution, $p(\eta(t_i) | y(t_i), \theta)$. The mean of the filtering distribution is plotted alongside previously unseen observations in Figure~\ref{fig:filteringDist}. It is also possible to draw from the parameter posterior when performing forecasting, in order to take into account uncertainty about the parameters, $\theta$.

\subsection{Irregular Observations: Temperature Sensors in the Urban Observatory}

One of the strengths of partially observed Markov process models with continuous time latent state is modelling irregularly observed data. The Urban Observatory Environment data~\cite{uo2014Temperature} contains many Temperature sensors, which vary in sampling rate and sometimes stop sampling altogether. Table~\ref{tbl:temper} shows the first six observations in July from a station outside the Key building in Newcastle Upon Tyne. The difference between most observations is approximately one minute, however the difference between observations four and five is three minutes. This is typical of sensor data, as some sensors sample adaptively, or lose power due to battery failure.

\begin{table}
\centering
  \begin{tabular}{|c|c|c|}
    \hline
    Timestamp           & Temperature (degree C) & $\Delta t$ \\
    \hline
    2016-07-01 00:00:41 &  10.9 & - \\
    2016-07-01 00:01:40 &  10.9 & 00:59 \\
    2016-07-01 00:02:43 &  10.9 & 01:03 \\
    2016-07-01 00:03:40 &  10.9 & 00:57 \\
    2016-07-01 00:06:40 &  10.9 & 03:00 \\
    2016-07-01 00:07:41 &  10.9 & 01:01 \\
    \vdots & \vdots & \vdots \\
    \hline
  \end{tabular}
\caption{The first 6 temperature measurements considered at the start of July, displaying the irregular arrival of observations from sensors. \label{tbl:temper}} 
\end{table}

\begin{figure}
\centering
\includegraphics[width=0.5\textwidth]{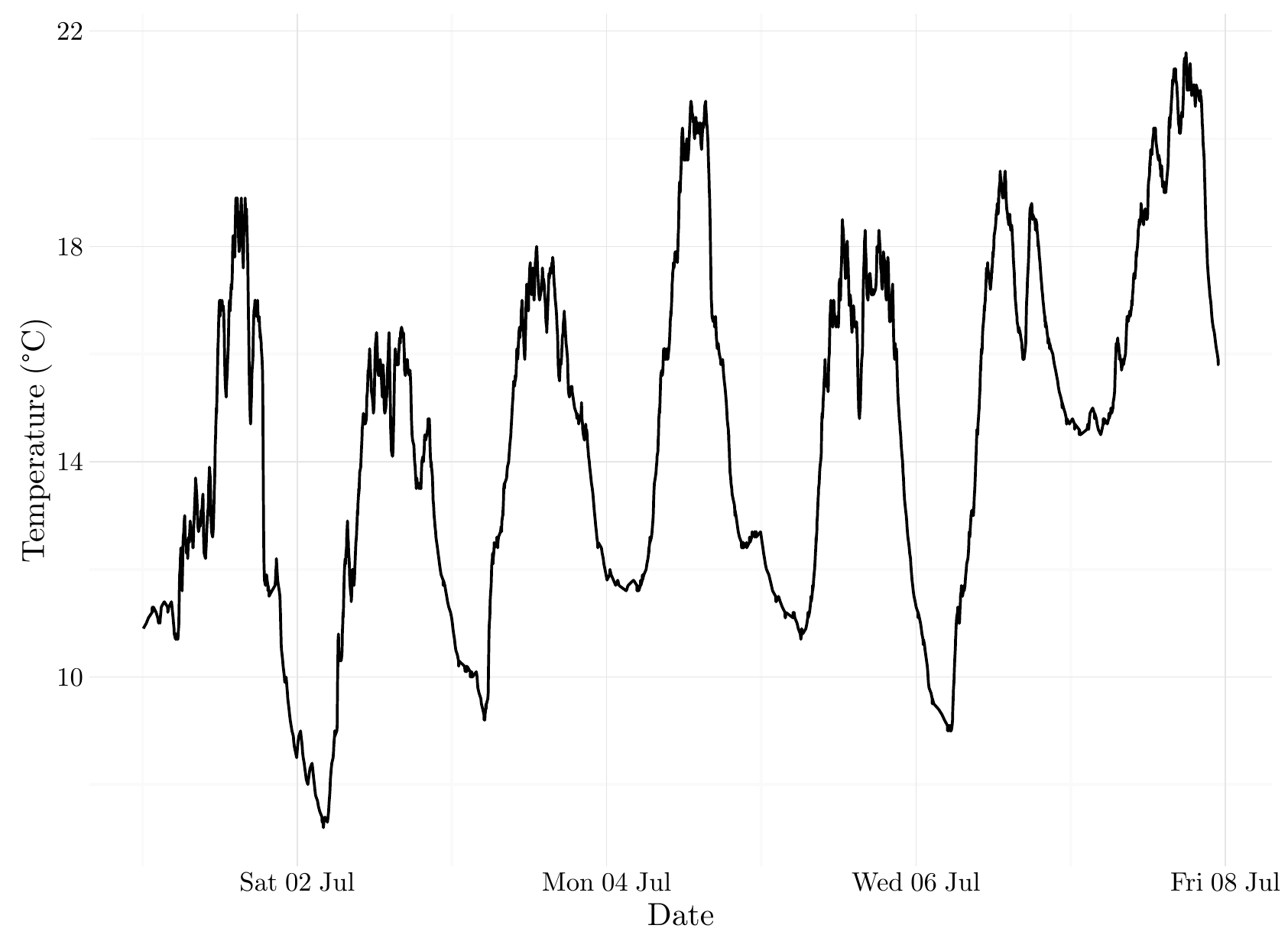}
\caption{Temperature recorded during the first week of July at the Key building in Newcastle Upon Tyne \label{fig:temperature}}
\end{figure}

Figure~\ref{fig:temperature} shows a plot of the temperature recorded in first week of July 2016. A suitable model to fit to this data is a seasonal model with a local-level trend. The first model in the composition is a linear Gaussian model with generalised Brownian motion representing the evolution of the latent state:

\begin{align*}
Y(t_i) &\sim \mathcal{N}(\mu(t_i), \sigma) \\
\mu(t_i) | \textbf{x}(t_i) &= \textbf{x}(t_i),  \\
\textrm{d}X(t) &= \mu X(t) + \sigma \textrm{d}W(t).
\end{align*}

\noindent
The linear transformation vector is the identity, $F^{(1)}_t = 1$ and the linking function is also the identity. The second model is a seasonal model with the Ornstein-Uhlenbeck process representing the evolution of the state space:

\begin{align*}
Y(t_i) &\sim \mathcal{N}(\mu(t_i), \sigma), \\
\mu(t) | \textbf{x}(t_i) &= F^{(2)T}_{t} \textbf{x}(t),  \\
\textrm{d}X(t) &= \boldsymbol{\alpha}(\boldsymbol{\theta} - \textrm{X}(t)) + \Sigma \textrm{d}W(t).
\end{align*}

\noindent
The linear transformation vector is used to represent the seasonality in the model and contains three harmonics:

\begin{equation*}
  F^{(2)}_t = \begin{pmatrix}
    \cos \omega t \\
    \sin \omega t \\
    \cos 2 \omega t \\
    \sin 2 \omega t \\
    \cos 3\omega t \\
    \sin 3\omega t
  \end{pmatrix}.
\end{equation*}

The model composition is then:

\begin{align*}
\label{eqn:tempmodel}
Y(t_i) &\sim \mathcal{N}(\mu(t_i), \sigma) \\
\mu(t_i) | \textbf{x}(t_i) &= F_{t_i}^T \textbf{x}(t_i)  \\
\textbf{X}(t_i) | \textbf{x}(t_{i-1}) &= \begin{pmatrix}
p_1(x^{(1)}(t_i)| x^{(1)}(t_{i-1}), \theta) \\
p_2(\textbf{x}^{(2)}(t_i)| \textbf{x}^{(2)}(t_{i-1}), \theta)
\end{pmatrix}.
\end{align*}

\noindent
The latent state is concatenated and the state from each model component evolves according to its transition kernel. The linking-equation is the identity, since the mean of the observation distribution (Gaussian) is free to vary over the entire real line. $F_t$ is a $(2*h + 1) \times 1$ vector, the seasonal model will have 3 harmonics, hence $F_t$ is a $7 \times 1$ vector:

\begin{equation*}
  F_t = \begin{pmatrix}
    1 \\
    \cos \omega t \\
    \sin \omega t \\
    \cos 2 \omega t \\
    \sin 2 \omega t \\
    \cos 3\omega t \\
    \sin 3\omega t
  \end{pmatrix}
\end{equation*}

\begin{figure}
\centering
\includegraphics[width=0.5\textwidth]{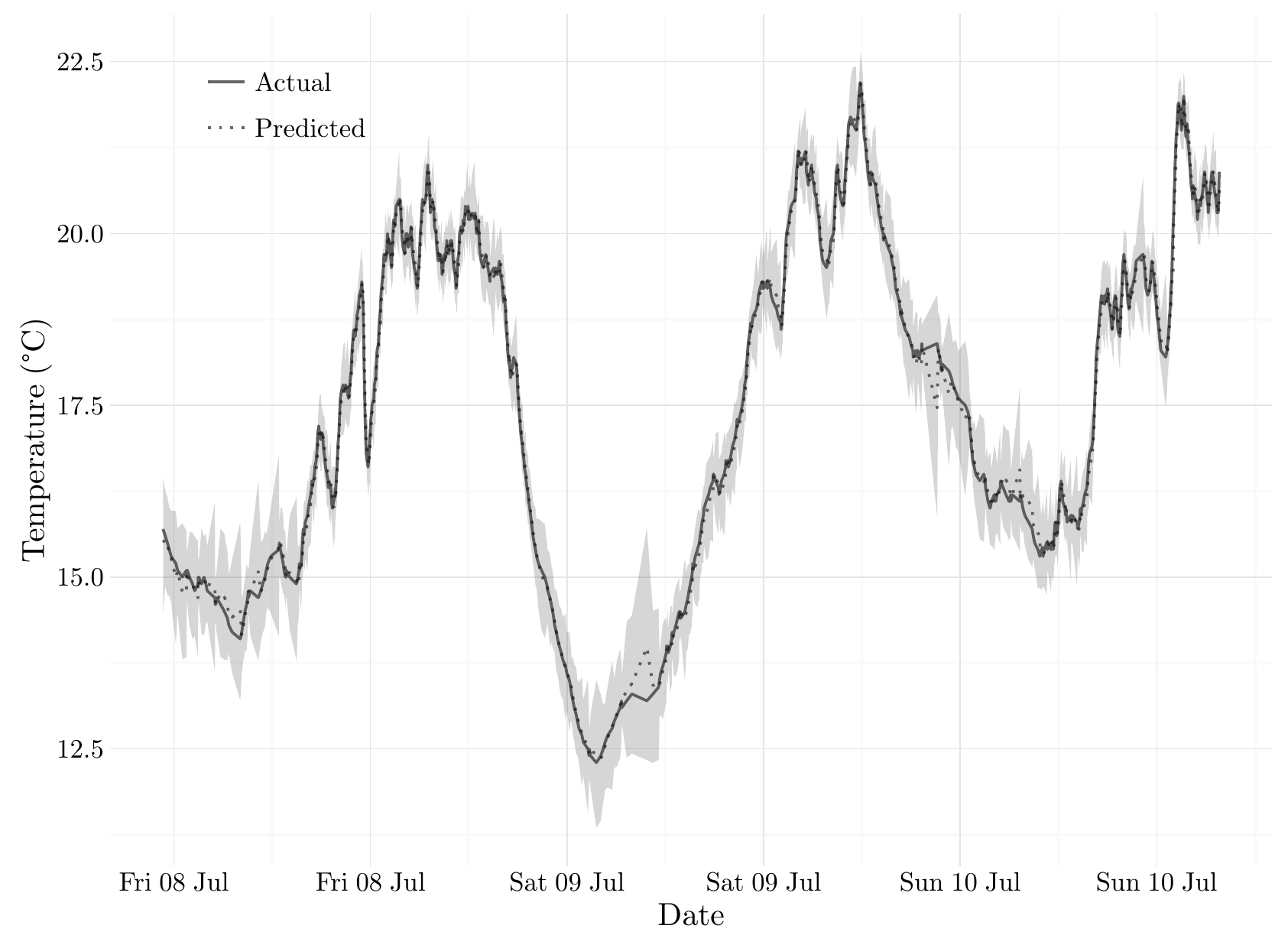}
\caption{Future observations of the temperature at the Key building in Newcastle Upon Tyne overlaid with a one-step forecast for the temperature model \label{fig:forecastTemperature}}
\end{figure}

This model shows the ability of the POMP models to accurately model irregularly observed data. Figure~\ref{fig:forecastTemperature} shows a plot of the one-step forecasts (stepped forward to the time of the next observation). The mean of the parameter posterior distribution was used to perform the one-step forecast.

\section{Conclusion}

Composable Markov process models have many applications. The class of composable models with the composition operator form a semi group, making it easy to build complex models by combining simple models. The use of the particle filter for simulation based inference has allowed for a flexible choice of observation models and a continuous time Markov process to represent the latent state. Further, by using a continuous time Markov process to represent the latent state, the requirement to simulate the state on a regular grid is relaxed. If there is an exact solution to the Markov process, then the latent state can be simulated forward to the next time of interest using the time increment and the previous realisation of the state. This allows for faster computation when using real world data, since many sensors sample at an adaptive rate, or consumption must be slowed down due to limited computing resources.

Incoming observations are represented as functional streams, which can represent live observations from a webservice or other source. This allows for a flexible and composable model framework which can be tested on simulated data and the same code deployed for real world applications. The particle filter is implemented as a functional \lstinline{fold}, a higher-order function allowing recursive data structures (such as the cons-list or cons-stream) to be reduced pairwise without side-effecting. The PMMH algorithm is written using \lstinline{iterate} (often called \lstinline{unfold}) which starts with a seed value and applies a function to it to create an infinite stream of iterations which can be manipulated using built-in stream processing functions. This allows for easy implementation of parallel chains, online-monitoring and constant-time memory usage, when used with asynchronous-IO.

The traffic and temperature sensors considered in the examples in Section~\ref{sec:example} are placed in close proximity to each other, hence they have highly correlated observations. The models presented in this paper currently consider the observations at each station as independent, but sharing information between sensors could mean more accurate inference. There is also a potential to determine faulty sensors, which do not record similar readings as those in its neighbourhood.

In some cases, results are needed quickly and accuracy is not of paramount concern. In this case, approximate inference could be performed using the Extended Kalman Filter and Unscented Kalman Filter mentioned in Section~\ref{sec:intro}.

Unusual events in the sensor network (for instance the presence of the world cup on traffic, pollution) can cause estimated parameters from past events to be ineffective to forecast future observations of the process. In this case it may be useful to have time varying parameters and estimate the parameters online.

All of the code used in this paper is available on GitHub~\footnote{https://git.io/statespace}.

\section*{Acknowledgements}

J. Law is supported by the Engineering and Physical Sciences Research Council, Centre for Doctoral Training in Cloud Computing for Big Data (grant number EP/L015358/1) and Digital Catapult 
Teaching Grant Award (KH153326).

\bibliography{composableModels}

\appendix

\include{Appendix}
\end{document}

%% file: Appendix.tex
\section{Appendix}

\subsection{Simulating Exponential Family Models}
\label{sec:sim}

In order to perform forecasting and interpolation, a method must be developed to forward simulate from POMP models. After determining the parameters for the model, as in section~\ref{sec:pmmh}, forward simulation of the model beyond observed values can be used to predict future observations along with the uncertainty of the predictions. In the case of the traffic data in section~\ref{sec:example}, forecasting can help with route planning and diversions. Similarly, simulating at periods of interest without real world observations is used to interpolate missing data.

In order to forward simulate from a POMP model, the following steps are required:

\begin{enumerate}
\item Generate a list of times to observe the process, $t_0, t_1 , \dots , t_n$
\item Initialise the state space at time $t_0$, by drawing from the initial distribution, $x(t_0) \sim p(x(t_0) | \theta)$
\item Calculate $\delta t = t_i - t_{i-1}$ and advance the state according to the Markov transition kernel $x_{t} \sim p(x(t_i) | x(t_{i-1}), \delta t, \theta)$
\item Apply the linear-deterministic transformation to the state, $\gamma(t_i) = f(t_i)(x(t_i))$ 
\item Transform the state into the parameter space of the observation model using the non-linear link-function, $\eta_t = g(f(t_i)(x(t_i)))$
\item Draw from the observation distribution, $y(t_i) \sim \pi(y(t_i) | x(t_i))$
\end{enumerate}

Forward simulation can also be useful for generating synthetic data to test inferential algorithms.

\subsection{Simulating the Log-Gaussian Cox-Process}
\label{sec:simlgcp}

As explained in Section~\ref{sec:sim}, simulating from POMP models is important for forecasting and interpolation. The steps uses to simulate from the LGCP are as follows, suppose the interval to be simulated on is $[0,T]$ and the last even occurred at time $t_0$:

\begin{enumerate}
  \item Simulate the log-Gaussian process, $\lambda(t)$ on a fine grid on the interval $[0,T]$
  \item $U_\lambda = \max\limits_{t \in [0,T]} \lambda(t)$
  \item Set $i = 1$
  \item Sample $t \sim Exp(U_\lambda)$
  \item Set $t_i = t_{i-1} + t$
  \item If $t_i > T$ stop
  \item Sample $u \sim U[0,1]$
  \item If $u \leq \lambda(t_i)/U_\lambda$ then accept $t_i$ as a new event time
  \item Set $i = i + 1$, go to 4.
\end{enumerate}

\subsection{Computing With Composed Models}
\label{sec:computingComposed}

The models are programmed using Scala, a language which runs on the JVM (Java Virtual Machine) and has good support for functional and concurrent programming, see~\cite{chiusano2014functional}. Functional programming is well suited to statistical inference problems and has been utilised in a similar context in a series of papers on the Kalman Filter,~\cite{Beckman2016}. These papers discuss Kalman folding, whereby a Kalman Filter is implemented on a collection of data using the catamorphism for lists and streams, (lazy lists) the fold.

A single model, $\mathcal{M}_i$, is given in equation~\ref{eqn:generalmodel}. First, consider a fully parameterised model. There are five functions which all models share in common and can be used to represent a single model in the Scala language:

\begin{enumerate}
\item A stochastic observation function: \\ \lstinline[language=scala]{val observation: Eta => Rand[Observation]}
\item A deterministic, non-linear linking function: \\ \lstinline[language=scala]{val link: Gamma => Eta}
\item A deterministic linear transformation function: \\ \lstinline[language=scala]{val f: (State, Time) => Gamma}
\item A stochastic function to advance the state: \\ \lstinline[language=scala]{val stepFun: (State, TimeIncrement) => Rand[State]}
\item A definition of the distribution of the initial state: \\ \lstinline[language=scala]{val x0: Rand[State]}
\item A likelihood function: \\ \lstinline[language=scala]{val dataLikelihood: (Eta, Observation) => LogLikelihood}
\end{enumerate}

The \lstinline{observation} method is a function from the transformed state $\eta(t)$ to a distribution over the observations $y$. The distribution is a monad and can be sampled from by calling the method \lstinline{draw}. The \lstinline[language=scala]{Rand} monad, along with many common discrete and continuous distributions are implemented in Scala Breeze\footnote{https://github.com/scalanlp/breeze}.

The linking function ensures the value of $\eta(t)$ is appropriate for the observation distribution. The Markov transition kernel of the state space is represented by (\lstinline{stepFun}) is a function from \lstinline[language=scala]{State} and \lstinline[language=scala]{TimeIncrement} to the next \lstinline[language=scala]{Rand[State]}. \lstinline[language=scala]{x0} returns a \lstinline[language=scala]{Rand} monad, with a \lstinline[language=scala]{draw} method, representing a random draw from the distribution of the initial state of the system. 

An important property of functional programming is referential transparency, this means that any function applied to a fixed set of arguments can be replaced by it's evaluation without affecting the final output of the program. This means purely functional programs are easier to reason about. Functions which produce a pseudo-random output are not pure and violate the principal of referential transparency. Hence, the observation function returns the \lstinline{Rand} monad which is manipulated via inference algorithms then sampled from at the last possible moment when the main program is run.

Probabilistic programming is an active area of research, \cite{scibior2015practical} have developed an approach to Bayesian Inference using monads in Haskell (a popular general purpose functional programming language). They develop a representation of a probability distribution as a Generalised Algebraic Data Type (GADT), then implement the particle MCMC algorithm Particle Independent Metropolis Hastings.

\subsubsection{A Binary Operation for Composing Models}

In order to compose models, define an associative binary operation called \lstinline[language=scala]{combine} which combines two models at a time. \lstinline[language=scala]{combine} accepts two un=parameterised models, (parameterisation is described in Section~\ref{sec:params}) \lstinline[language=scala]{m1} and \lstinline[language=scala]{m2} and returns a third model representing the sum of the two models. In this way, the models form a semi-group, since the combine operation is associative and closed.

Further, an identity model, $e$, can be defined is defined such that for any model $\mathcal{M}_a$, $e \star \mathcal{M}_a = \mathcal{M}_a = \mathcal{M}_a \star e$. Un-parameterised models now form a monoid. A monoid is an algebraic data type commonly found in functional programming.

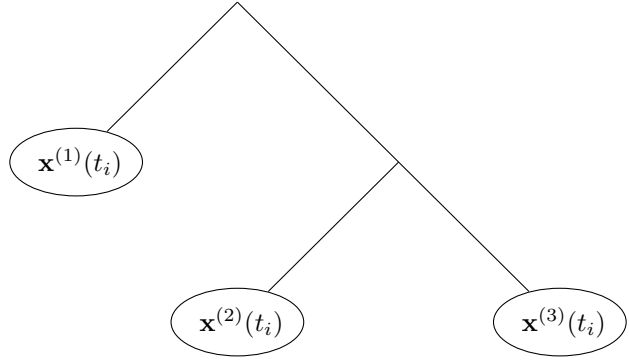
\begin{figure}

\tikzstyle{line} = [draw, -]
\tikzstyle{state} = [draw, ellipse, text width=1cm, align=center]

\begin{center}
  \begin{tikzpicture}[node distance = 2cm, auto]
    \coordinate (1) {};
    \node [state, below left of=1, node distance=3cm] (2) {$\textbf{x}^{(1)}(t_i)$};
    \coordinate [below right of=1, node distance=3cm] (3) {};
    \node [state, below left of=3, node distance=3cm] (4) {$\textbf{x}^{(2)}(t_i)$};
    \node [state, below right of=3, node distance=3cm] (5) {$\textbf{x}^{(3)}(t_i)$};
    
    \path [line] (1) -- (2);
    \path [line] (1) -- (3);
    \path [line] (3) -- (4);
    \path [line] (3) -- (5);
    
  \end{tikzpicture}
\end{center}

\caption{State Space as a Binary Tree, this composed model state consists of the combined states of three models\label{fig:binTree}}
\end{figure}

In order to define the \lstinline{combine} operation, firstly consider combining the state space of two models: A binary tree is used in order to represent the state space of a model, this is depicted in figure~\ref{fig:binTree}. For a single model, the state space is called a \lstinline[language=scala]{LeafState}. In order to represent the state of two models, the state of each model is combined into a \lstinline{BranchState} with the left and right branch corresponding to the \lstinline{LeafState} of each model. A binary tree for the state is defined in Scala as:

\begin{lstlisting}
sealed trait State
case class LeafState(x: Vector[Double]) extends State
case class BranchState(left: State, right: State) extends State
\end{lstlisting}

Since the \lstinline{LeafState} and \lstinline{BranchState} both extend the \lstinline{State}, when a function accepts a parameter of type \lstinline{State} it can be either a \lstinline{LeafState} or a \lstinline{BranchState}. Pattern matching is used in order to decompose the state trees and perform functions on the values contained in the leaf nodes.

Now, consider advancing the state space of a combined model, by drawing from the transition kernel conditional on the current state using \lstinline[language=scala]{stepFun}. Each model in the combined model has a corresponding \lstinline{LeafState} and step function, each step function must act on the corresponding state. To advance the state of a composed model of two, we act on the left-hand (correspondingly right-hand) model's state space with the left-hand (right-hand) model's step function.

\begin{lstlisting}
val stepFunction = (s: State, dt: TimeIncrement) => s match {
      case BranchState(ls, rs) =>
        for {
          l <- mod1.stepFunction(ls, dt)
          r <- mod2.stepFunction(rs, dt)
        } yield BranchState(l, r)
\end{lstlisting}

The application of the vector $F_t$ is performed using the function \lstinline[language=scala]{def f: (State, Time) => Observation}. For two models, this function expects a \lstinline{BranchState}, which can be decomposed using pattern matching:

\begin{lstlisting}
val f = (s: State, t: Time) => s match {
      case BranchState(leftState, rightState) =>
        mod1.f(leftState, t) + mod2.f(rightState, t)
    }
\end{lstlisting}

The observation function is taken to be that of the left-hand model, \lstinline[language=scala]{def observation = eta => mod1.observation(eta)}. Which means the linking function must also be that of the left-hand model \lstinline[language=scala]{def link = x => mod1.link(x)}.

In order to simulate a model or initialise a particle filter, there must be an initial distribution for the state, this is given by the \lstinline{x0} function. Adding two models will result in a \lstinline{BranchState} for the initial state:

\begin{lstlisting}
val x0 = for {
  l <- mod1.x0
  r <- mod2.x0
} yield BranchState(l, r)
\end{lstlisting}

\subsubsection{Parameterising the Model}
\label{sec:params}

The parameters of a model given in Equation~\ref{eqn:genpomp}, are that of the Markov transition for the state space, $\theta$ in $p(\textbf{x}(t_i)|\textbf{x}(t_{i-1}), \theta)$, the parameters of the initial system state $\textbf{x}(0) \sim p(\textbf{x}(0) | \theta)$ and any additional parameters for the observation function. An unparameterised model is defined as a function from \lstinline{Parameters => Model}, which can then be supplied with appropriate parameters to form a Model.

When composing two models, a set of parameters for each model must be combined into a new set of parameters for the new model. As with the state-space, it is natural to model the parameters of a composed model as a binary tree. A single model is parameterised by a \lstinline[language=scala]{LeafParameter}. When composing two models, leaf parameters will be combined into a branch, termed \lstinline[language=scala]{BranchParameter}. Branch parameters have a left and a right branch, in the case of a composition of two models, each branch corresponds to a single \lstinline[language=scala]{LeafParameter} object. Parameters must be combined in the same order as models are combined, ie. combining parameters is not commutative. There is an associative infix operator, \lstinline[language=scala]{|+|}, for creating \lstinline[language=scala]{BranchParameter} objects:

\begin{lstlisting}
val (p, p1, p2) = (LeafParameter(.), LeafParameter(.), LeafParameter(.))
val combParams = p |+| p1 |+| p2
\end{lstlisting}

This is equivalent to constructing the parameters as nested branch parameters:

\begin{lstlisting}
val combParams = BranchParameter(BranchParameter(p, p1), p2)
\end{lstlisting}

Consider a composed model consisting of two single models, this means the state will be a \lstinline{BranchState} containing two \lstinline{LeafState}s. The Parameters of the model will be a \lstinline{BranchParameter} containing two \lstinline{LeafParameter}s. In order to advance the state pattern matching is used to decompose both the parameters and the state simultaneously:

\begin{lstlisting}
def stepFunction = (s: State, dt: TimeIncrement) => (s: State, p: Parameters) match {
    case (BranchState(ls, rs), BranchParameter(lp, rp)) =>
        for {
          l <- mod1(lp).stepFunction(ls, dt)
          r <- mod2(rp).stepFunction(rs, dt)
        } yield BranchState(l, r)
    }
\end{lstlisting}